\documentclass[twocolumn]{aastex63}

\newcommand\teff{\mbox{$T_\mathrm{eff}$}}
\newcommand\vtan{\mbox{$V_\mathrm{tan}$}}
\newcommand\kms{{km s$^{-1}$}}
\usepackage{color}

\begin{document}

\title{Astrometry and Photometry for $\approx$1000 L, T, and Y Dwarfs from the UKIRT Hemisphere Survey}

\correspondingauthor{Adam C. Schneider}
\email{adam.c.schneider4.civ@us.navy.mil}

\author[0000-0002-6294-5937]{Adam C. Schneider}
\affil{United States Naval Observatory, Flagstaff Station, 10391 West Naval Observatory Rd., Flagstaff, AZ 86005, USA}

\author[0000-0002-4603-4834]{Jeffrey A. Munn}
\affil{United States Naval Observatory, Flagstaff Station, 10391 West Naval Observatory Rd., Flagstaff, AZ 86005, USA}

\author{Frederick J. Vrba}
\affil{United States Naval Observatory, Flagstaff Station, 10391 West Naval Observatory Rd., Flagstaff, AZ 86005, USA}

\author[0000-0002-3858-1205]{Justice Bruursema}
\affil{United States Naval Observatory, Flagstaff Station, 10391 West Naval Observatory Rd., Flagstaff, AZ 86005, USA}

\author[0000-0002-2968-2418]{Scott E. Dahm}
\affil{Gemini Observatory/NSF’s NOIRLab, 950 N. Cherry Avenue, Tucson, AZ, 85719, USA}

\author[0000-0002-3858-1205]{Stephen J. Williams}
\affil{United States Naval Observatory, Flagstaff Station, 10391 West Naval Observatory Rd., Flagstaff, AZ 86005, USA}

\author[0000-0003-2232-7664]{Michael C. Liu}
\affil{Institute for Astronomy, University of Hawaii at Manoa, Honolulu, HI 96822, USA}

\author[0000-0002-5604-5254]{Bryan N. Dorland}
\affil{United States Naval Observatory, 3450 Massachusetts Ave NW, Washington, DC 20392-5420, USA}

\begin{abstract}

We present positions, proper motions, and near-infrared photometry for 966 known objects with spectral types later than M observed as part of the the UKIRT Hemisphere Survey (UHS).  We augment the photometry and astrometry from UHS with information from Gaia DR3, Pan-STARRS DR2, and CatWISE 2020 to produce a database of homogeneous photometry and astrometry for this sample. The multi-epoch survey strategy of UHS allows us to determine proper motions for most sources, with a median proper motion uncertainty of $\sim$3.6 mas yr$^{-1}$. Our UHS proper motion measurements are generally in good agreement with those from Gaia DR3, Pan-STARRS, and CatWISE 2020, with UHS proper motions typically more precise than those from CatWISE 2020 and Pan-STARRS but not Gaia DR3.  We critically analyze publicly available spectra for 406 members of this sample and provide updated near-infrared spectral types for $\sim$100 objects.  We determine typical colors as a function of spectral type and provide absolute magnitude vs.\ spectral type relations for UHS $J$- and $K$-band photometry.  Using newly determined proper motions, we highlight several objects of interest, such as objects with large tangential velocities,  widely separated co-moving companions, and potential members of young nearby associations.  
\end{abstract}


\section{Introduction}
\label{sec:intro}

Large-scale optical and infrared surveys, such as Deep Near-Infrared Southern Sky Survey (DENIS; \citealt{epchtein1997}), the Sloan Digital Sky Survey (SDSS; \citealt{york2000}), the Two Micron All-Sky Survey (2MASS; \citealt{skrutskie2006}), the Wide-field Infrared Survey Explorer (WISE; \citealt{wright2010}) and the Panoramic Survey Telescope And Rapid Response System (Pan-STARRS; \citealt{kaiser2010}), have been instrumental in the discovery and characterization of the vast majority of the known ultracool population.  These efforts, combined with dedicated spectroscopic follow-up campaigns, have led to the definition of the L, T, and Y spectral classes (e.g., \citealt{kirkpatrick2005, cushing2011}), as well as sub-populations within these classes, such as low-metallicity subdwarfs (e.g., \citealt{burgasser2003b, zhang2017}) and young, low-gravity substellar objects (e.g., \citealt{kirkpatrick2008, allers2013, faherty2016}).    

Large, homogeneous samples of brown dwarfs have been valuable for characterizing the known population as well as  analyzing individual objects (e.g., \citealt{kirkpatrick2011, best2018}).  The UKIRT Hemisphere Survey (UHS) covers approximately 12,700 deg$^2$ of the northern hemisphere with near-infrared $J$ and $K$ filters between 0$\degr$ and 60$\degr$ \citep{dye2018, bruursema2023}.  We have gathered photometric and astrometric information for all known L, T, and Y dwarfs within the UHS footprint, using both the previously released $J$-band portion of the survey (DR1; \citealt{dye2018}) and newly released $K$-band UHS data (DR2; \citealt{bruursema2023}). The creation of the sample is described in Section \ref{sec:sample}.  We describe the photometry of the sample in Section \ref{sec:phot} and describe new proper motion measurements in Section \ref{sec:pms}.  In Section \ref{sec:ooi} we describe efforts to identify new young moving group members, objects with large tangential velocities, and co-moving companions amongst this sample.  We summarize our results in Section \ref{sec:summary}.

\section{Sample Selection}
\label{sec:sample}

The goal of this work is to compile UHS astrometry and photometry for known L, T, and Y dwarfs in the UHS footprint.  We select only those objects with spectroscopically determined spectral types (i.e., no photometric brown dwarf candidates). Our initial list of candidates for the creation of this sample was the UltracoolSheet\footnote{http://bit.ly/UltracoolSheet}, a public compilation of ultracool dwarfs (e.g., spectral types later than $\sim$M6) and directly imaged exoplanets compiled in \cite{best2020}.  To this list, we added several papers with spectroscopically classified L, T, or Y dwarfs that were either not included \citep{robert2016, kellogg2017, kuchner2017, perez2017, schneider2017, perez2018, greco2019, kiman2019, zhang2019} or published after the creation of the UltracoolSheet in 2020 \citep{faherty2020, meisner2020, schneider2020, zhang2020, faherty2021, marocco2021, meisner2021, kirkpatrick2021, kiwy2022, schneider2022, softich2022, vos2022}. 

\begin{figure}
\plotone{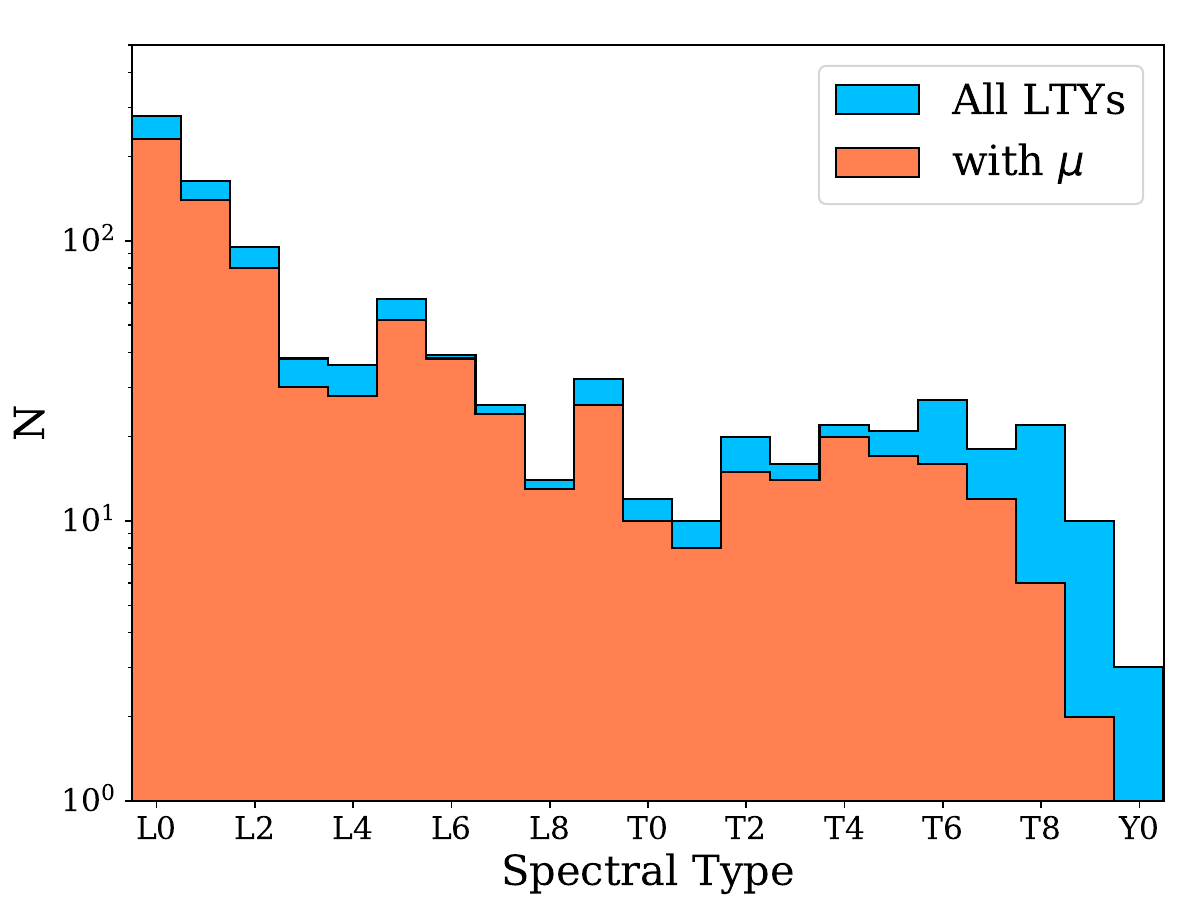}
\caption{Histogram of spectral types for the sample of known L, T, and Y dwarfs in the UHS catalog.  Note the logarithmic y-scale.  The orange histogram shows the subset of objects from the sample that have newly measured proper motions in this work. The entire sample contains 966 L, T, and Y dwarfs, of which 782 have newly measured proper motions using UHS data. }  
\label{fig:plot1}
\end{figure}

Because L, T, and Y dwarfs can have large proper motions, we first performed a cross-match with a large search radius (1\arcmin) to determine which objects from this initial list were likely covered by the UHS survey.  Then we individually checked for detections of each object using UHS images and previously measured positions.  A total of 966 L, T, and Y dwarfs were found to have at least one detection in UHS DR2.  Figure \ref{fig:plot1} shows a histogram of the spectral types of this sample (see Section \ref{sec:spts}).  As seen in the figure, a significant portion of this sample consists of early L dwarfs from various Sloan Digital Sky Survey (SDSS) surveys \citep{west2008, schmidt2010, west2011, kiman2019}, though every spectral type bin earlier than Y0 has $\geq$10 objects.  The histogram also shows the portion of the entire sample for which new proper motion measurements were calculated using UHS data ($\sim$81\% of all objects), discussed further in Section \ref{sec:pms}.   

Positions and UHS $J$- and $K$-band photometry for objects in this sample pulled directly from UHS DR2 \citep{bruursema2023} are given in Table \ref{tab:props}.  We also include $J$- and $K$-band positions, uncertainties, and epochs for each object after re-registering to the Gaia DR3 \citep{gaia2022} reference frame (see Section \ref{sec:pms}). 

\subsection{Additional Photometry and Astrometry}
We include photometry and astrometry from Gaia DR3 \citep{gaia2022}, Pan-STARRS DR2 \citep{chambers2016, magnier2020}, and CatWISE 2020 \citep{marocco2021} for each object, when available.  These data allow us to evaluate our derived proper motions and to compile more complete spectral energy distributions in order to determine color trends with spectral type utilizing UHS photometry.  Relevant photometry and astrometric measurements from these surveys are given in Table \ref{tab:props}.

\subsection{Spectral Types}
\label{sec:spts}

Spectral types were taken from the literature for most objects, with exceptions detailed in the following paragraphs.  For objects with multiple spectral type determinations, we use the first measured type if all measurements agree, and preferentially use the most recent measurements when disagreements arise.  We list optical and near-infrared spectral types for this sample in Table \ref{tab:props}.  

In an effort to have consistent spectral types, we searched for existing IRTF/SpeX near-infrared spectroscopy in the SpeX Prism Library Analysis Toolkit (SPLAT; \citealt{burgasser2017}) for every object in this sample to confirm or refine spectral types.  Four-hundred and six objects were found to have near-infrared spectra in the SPLAT archive.  To derive near-infrared spectral types, we compared to the T dwarf spectral standards in \cite{burgasser2006} and L dwarf spectral standards of \cite{kirkpatrick2010}, with the exception that we use 2MASSI J0825196$+$211552 as the L7 standard as recommended in \cite{cruz2018}. We also use the subdwarf standards defined in \cite{greco2019} when appropriate.  Using such existing spectra, we generally find good agreement with previously determined spectral types to within 0.5 subtypes.  We found 48 objects with previously determined near-infrared spectral types that differ by a full subtype or more from our determination, which are listed in Table \ref{tab:spts1}. 

\startlongtable
\begin{deluxetable}{ccccccccc}
\label{tab:spts1}
\tabletypesize{\scriptsize}
\tablecaption{Updated Near-Infrared Spectral Types }
\tablehead{
\colhead{Name} & \colhead{Prev.~NIR} & \colhead{Ref.} & \colhead{New} \\
\colhead{          } & \colhead{Type} & \colhead{      } & \colhead{Type} }
\startdata
WISEA J000627.85$+$185728.8 & L7 & 1 & L6 (sl. red) \\
2MASSW J0015447$+$351603 & L1 & 2 & L2 \\
2MASS J00282091$+$2249050 & L5 & 3 & L6 \\
2MASSW J0030438$+$313932 & L3.2 & 2 & L2 \\
2MASSW J0228110$+$253738 & L0 & 6 & L1 \\
SDSS J024256.98$+$212319.6 & L4 & 7 & L5.5 \\
2MASS J03250136$+$2253039 & L3.3 & 2 & L2 \\
2MASSJ03440892$+$0111251 & L2.9: & 2 & L1 (sl. blue) \\
2MASS J07244848$+$2506143 & L4 & 9 & L3 \\
2MASSW J0740096$+$321203 & L4 & 2 & L5 \\
SDSS J080048.13$+$465825.5 & L1.3 & 2 & L0.5 \\
SDSS J081253.19$+$372104.2 & L0.5 pec & 10 & M9 (sl. red) \\
SDSS J081757.49$+$182405.0 & L2 & 2 & L0.5 \\
SDSS J084333.28$+$102443.5 & L2.7: & 2 & L1 (sl. blue) \\
SDSS J092308.70$+$234013.7 & L2.3 & 2 & L1 (sl. blue) \\
2MASS J09325053$+$1836485 & L6 & 9 & L4.5 \\
SDSS J094047.88$+$294653.0 & L2: & 10 & L1 \\
2MASS J09481259$+$5300387 & L2 & 9 & L1 \\
2MASS J10271549$+$5445175\tablenotemark{a} & L7 & 9 & L2 VL-G \\
SDSS J103321.92$+$400549.5 & L6 & 7 & L5 (blue) \\
SDSS J111320.16$+$343057.9 & L3 & 7 & L2 \\
2MASS J12312141$+$4959234 & L3.4 & 2 & L2 \\
2MASS J12352675$+$4124310 & L5 & 9 & L2: (red) \\
2MASS J12453705$+$4028456 & L1 & 9 & L2 (blue) \\
2MASSW J1246467$+$402715 & L4 & 2 & L5 \\
2MASSI J1305410$+$204639 & L6.5 & 2 & L5 \\
2MASS J13451417$+$4757231 & L3 & 9 & L5.5 \\
SDSS J142612.86$+$313039.4 & L4:: & 2 & L5 \\
SDSS J143832.63$+$572216.9 & L4.6 & 2 & L3.5 \\
SDSS J152039.82$+$354619.8 & T0$\pm$1 & 7 & L9 \\
2MASS J15311344$+$1641282 & L1 & 12 & L2 \\
SDSS J154849.02$+$172235.4 & L5 & 7 & L8.5 \\
2MASS J15500191$+$4500451 & L6 & 9 & L2 (red) \\
SDSS J155120.86$+$432930.3 & L3.1 & 2 & L2 \\
2MASS J15543602$+$2724487 & L5 & 9 & L6 \\ 
2MASS 16094569$+$1426422 & L4 & 9 & L2 \\
SDSS J161459.98$+$400435.1 & L2 & 13 & L3 \\
SIMP J16270845$+$0546304 & L0 & 10 & M9 \\
2MASS J16304139$+$0938446 & L0.4: & 2 & L1.5 \\
2MASS J17120142$+$3108217 & L3 & 9 & L2 \\
2MASS J17161258$+$4125143 & L4 & 14 & L2 \\
2MASSI J1721039$+$334415 & L5.3: & 2 & L3 (blue) \\
LSPM J1731$+$2721 & L0 & 15 & M9 \\
WISE J173332.50$+$314458.3 & L2 & 13 & L3.5 (red) \\
WISEA J174336.62$+$154901.3 & L1 pec (blue) & 8 & M9.5 \\
2MASSI J2057153$+$171515 & M9.9 & 2 & L1 (sl. red) \\
SDSS J214046.55$+$011259.7 & L4.5 & 3 & L2 (sl. blue) \\
PSO J344.8146$+$20.1917 & L2.5 & 16 & L4 (sl. red) \\
\enddata
\tablenotetext{a}{See text in Section \ref{sec:sample} for a discussion of this object.}
\tablerefs{(1) \cite{schneider2016}; (2) \cite{bardalez2014}; (3) \cite{burgasser2010}; (4) \cite{burgasser2004}; (5) \cite{kirkpatrick2011}; (6) \cite{wilson2003}; (7) \cite{chiu2006}; (8) \cite{luhman2014b}; (9) \cite{kellogg2017}; (10) \cite{robert2016};  (11) \cite{sheppard2009}; (12) \cite{faherty2009}; (13) \cite{thompson2013}; (14) \cite{kellogg2015}; (15) \cite{allers2013}; (16) \cite{best2015} }
\end{deluxetable}

In this work, we prefer near-infrared spectral types over optical spectral types because of the infrared nature of the UHS survey.  Thus, for objects with only optical types in the literature, we again searched the SPLAT database in order to update to near-infrared types when possible.  New near-infrared types were found for eleven objects with existing optical spectral types and are listed in Table \ref{tab:spts2}. One object, (SDSS J153012.87$+$514717.1) is typed as L0 in the optical \citep{west2008}, but M9 in the near-infrared, and is thus excluded from the final sample.  

\subsection{Notes on Object Spectral Types}

\subsubsection{2MASS J10271549$+$5445175}

\begin{figure}
\plotone{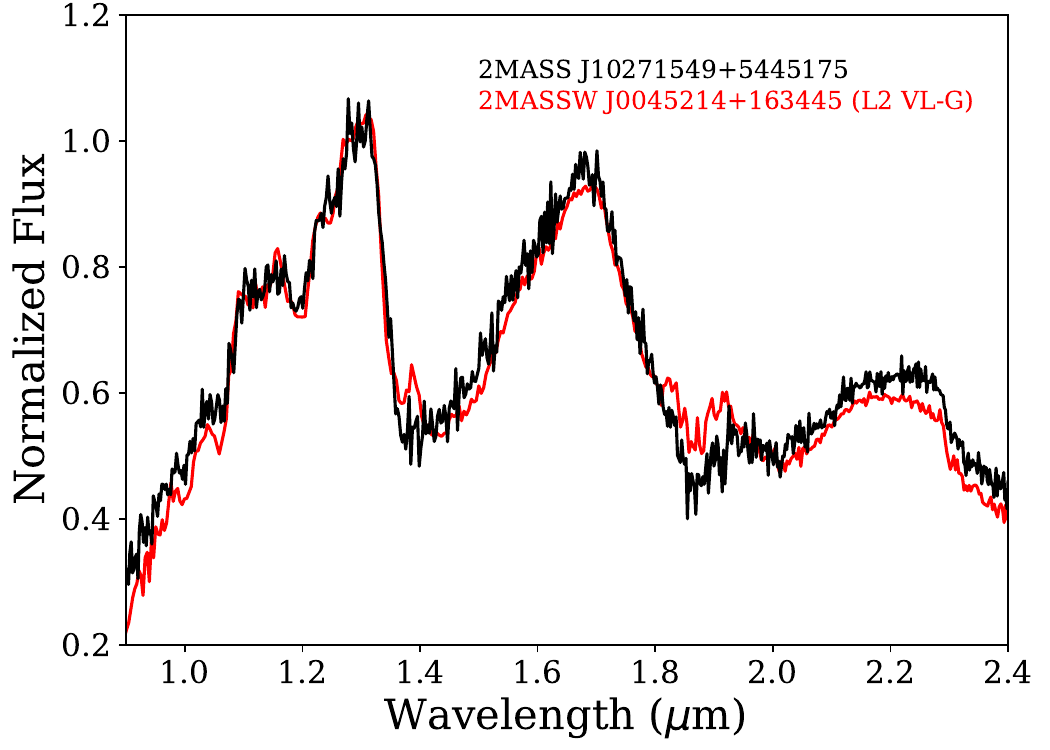}
\caption{The near-infrared spectrum of 2MASS J10271549$+$5445175 compared to the L2 VL-G object 2MASSW J0045214$+$163445.  Each spectrum is normalized to the $J$-band peak between 1.27 and 1.29 $\mu$m.  }  
\label{fig:J1027}
\end{figure}

2MASS J10271549$+$5445175 was discovered in \cite{kellogg2017} and assigned a spectral type of L7.  Our reanalysis using the available spectrum in SPLAT instead found that this object is likely a low-gravity early type L dwarf.  Figure \ref{fig:J1027} shows a comparison of the near-infrared spectrum of this object to 2MASSW J0045214$+$163445, which is typed as L2 VL-G in \cite{allers2013} and L2$\gamma$ in \cite{faherty2016}.  We find that the spectrum of 2MASS J10271549$+$5445175 is an excellent match to that of 2MASSW J0045214$+$163445, and we therefore assign this object a spectral type of L2 VL-G (or L2$\gamma$).  This classification is bolstered by the low-gravity indices defined in \cite{allers2013} applicable to the low-resolution spectrum -- we find FeH$_z$, VO$_z$, K I$_J$ and $H-$cont values of 1.052, 1.079, 0.988, and 0.989, respectively, which leads to a gravity score of 2022 (i.e., VL-G).  This object is discussed further in Section \ref{sec:arg}.     

\begin{deluxetable}{ccccccccc}
\label{tab:spts2}
\tablecaption{New Near-Infrared Spectral Types }
\tablehead{
\colhead{Name} & \colhead{Opt.} & \colhead{Ref.} & \colhead{NIR} \\
\colhead{          } & \colhead{Type} & \colhead{      } & \colhead{Type} }
\startdata
2MASSI J0025036$+$475919 & L4: & 1 & L3.5 (red) & \\
2MASSI J0213288$+$444445 & L1.5 & 2 & L0.5 \\ 
SDSS J080027.57$+$551134.1 & L1 & 3 & L1 \\
2MASSI J1117369$+$360936 & L0? & 2 & L1 \\
SDSS J134148.85$+$551046.2 & L2 & 4 & L3 \\
SDSS J153012.87$+$514717.1 & L0 & 5 & M9 \\
SDSS J161611.36$+$521328.0 & L0 & 5 & L0 \\
SDSS J163437.19$+$233620.5 & L1 & 4 & L1 \\
2MASSW J1841086$+$311727 & L4 pec & 6 & L5 \\
2MASS J21522609$+$0937575 & L7: & 7 & L7 \\
2MASS J22490917$+$3205489 & L5 & 1 & L5 (red) \\
\enddata
\tablerefs{(1) \cite{cruz2007}; (2) \cite{cruz2003}; (3) \cite{schmidt2010}; (4) \cite{kiman2019}; (5) \cite{west2008}; (6) \cite{kirkpatrick2000}; (7) \cite{cruz2018}  }
\end{deluxetable}

\subsubsection{Newly Classified Subdwarfs}
The near-infrared spectra of several objects showed hallmarks of being low-metallicity subdwarfs, such as suppressed emission in the $H-$ and $K$-bands due to enhanced collisionally-induced H$_2$ absorption (see e.g., \citealt{zhang2017}).  To classify these objects, we compared their spectra to the near-infrared subdwarf standards suggested in \cite{greco2019}.  Since there is no sdL0 standard in \cite{greco2019}, we propose the subdwarf SSSPM J1444$-$2019 \citep{scholz2004} as the sdL0 near-infrared standard, as it is classified as an sdL0 in the optical and near-infrared in \cite{kirkpatrick2016}.  Another object, 2MASS J00412179$+$3547133, was given a near-infrared classification of sdL? in \cite{burgasser2004} and classified in the optical as sdL0.5 \citep{zhang2017}.  The spectrum of this object is intermediate between our sdL0 standard (SSSPM J1444$-$2019) and the sdL1 standard (2MASS J17561080$+$2815238; \citealt{kirkpatrick2010}).  We therefore suggest 2MASS J00412179$+$3547133 as the sdL0.5 near-infrared template.  

\begin{figure*}
\plotone{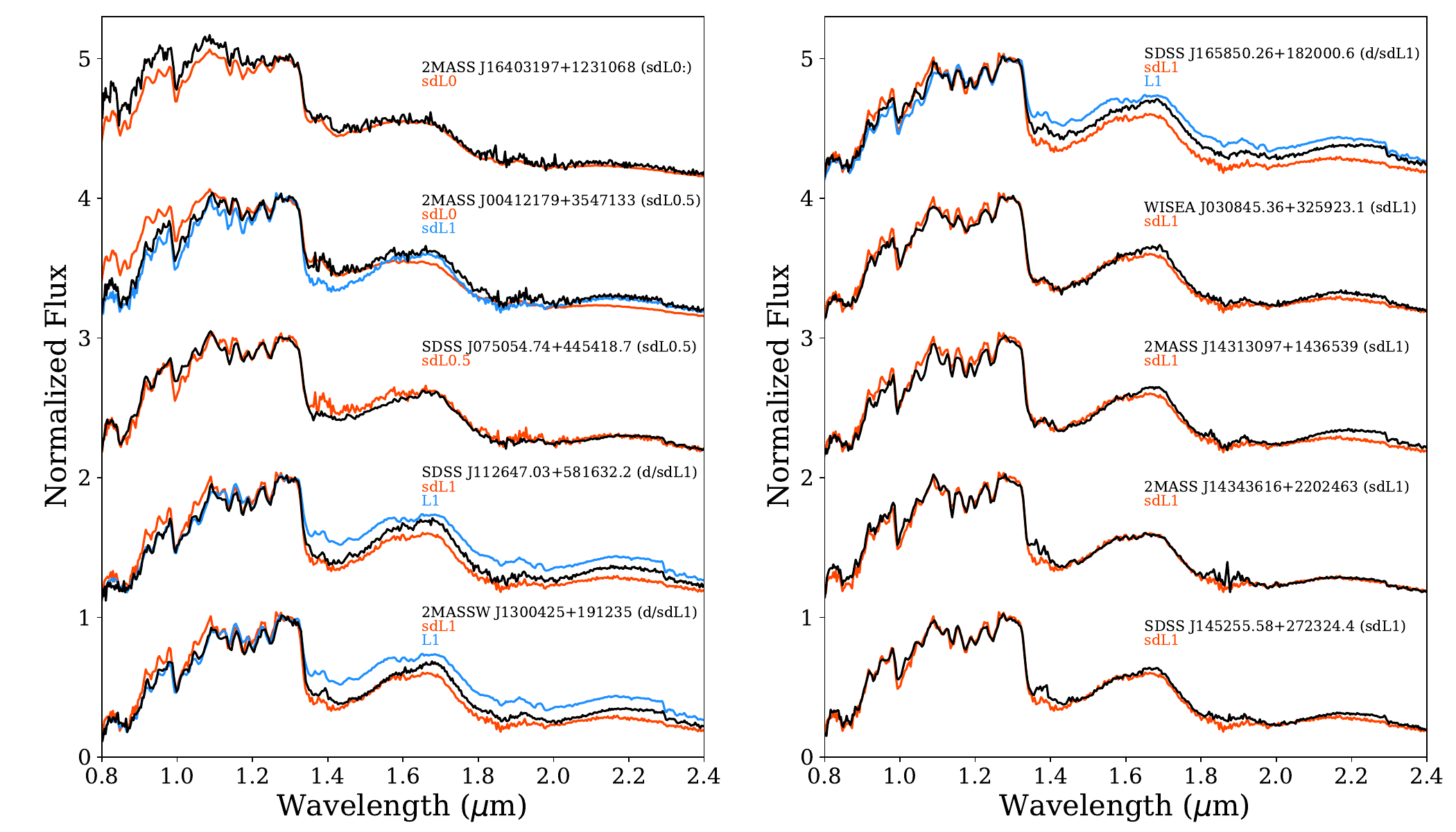}
\caption{Objects with new (or updated) sdL classifications.  Each spectrum is normalized to the $J$-band peak between 1.27 and 1.29 $\mu$m and offset by integer numbers for clarity.  Spectral standards are plotted as orange or blue with types labeled in the figure.}  
\label{fig:sds}
\end{figure*}

Objects with new or updated spectral types that place them in the sdL class are listed in Table \ref{tab:sdspts} and shown in Figure \ref{fig:sds}.  Note that three objects (SDSS J112647.03$+$581632.2, 2MASSW J1300425$+$191235, and SDSS J165850.26$+$182000.6) are classified as d/sdL1 and are compared to the sdL1 and L1 spectral standards in the figure.  Several of these objects have kinematics consistent with membership in the old, thick disk population, and are discussed further in Section \ref{sec:sds}.

\begin{deluxetable}{ccccrrr}
\label{tab:sdspts}
\tablecaption{Newly Classified Subdwarfs }
\tablehead{
\colhead{Name} & \colhead{Prev.~NIR} & \colhead{Ref.} & \colhead{New} \\ 
\colhead{          } & \colhead{Type} & \colhead{      } & \colhead{Type}}
\startdata
2MASS J00412179$+$3547133\tablenotemark{a} & sdL? & 1 & sdL0.5 \\ 
WISEA J030845.36$+$325923.1 & L1 pec (blue) & 2 & sdL1 \\ 
SDSS J075054.74$+$445418.7 & M8 pec & 3 & sdL0.5 \\ 
SDSS J112647.03$+$581632.2 & L1.3 & 4 & d/sdL1 \\ 
2MASSW J1300425$+$191235 & L1.7 & 4 & d/sdL1 \\ 
2MASS J14313097$+$1436539 & L2 & 5 & sdL1 \\ 
2MASS J14343616$+$2202463 & sdM9 & 5 & sdL1 \\ 
SDSS J145255.58$+$272324.4 & L0.2: & 4 & sdL1 \\ 
2MASS J16403197$+$1231068 & sdM8? & 1 & sdL0: \\ 
SDSS J165850.26$+$182000.6 & L0.9 & 4 & d/sdL1 \\ 
\enddata
\tablenotetext{a}{Classified as sdL0.5 in the optical by \cite{zhang2017}, and suggested in this work as the sdL0.5 near-infrared template.}
\tablerefs{(1) \cite{burgasser2004}; (2) \cite{luhman2014b}; (3) \cite{thompson2013}; (4) \cite{bardalez2014}; (5) \cite{sheppard2009}}
\end{deluxetable}

\section{Photometry}
\label{sec:phot}

Photometry from the UHS survey is based on the UKIRT photometric system \citep{hodgkin2009}, which was designed to closely match photometry from the Mauna Kea Observatories (MKO) system \citep{simons2002}.  $J$- and $K$-band photometry for all known L, T, and Y dwarfs detected in the UHS survey is taken directly from UHS DR2 and given in Table \ref{tab:props}.  

This photometry can be used to find typical UHS colors and absolute magnitudes for the various spectral types included in this sample.  There are many factors that determine a specific L, T, or Y dwarf color, most notably effective temperature (\teff), but also surface gravity (e.g., \citealt{allers2013}), inclination angle (e.g., \citealt{vos2017}), unresolved binarity (e.g., \citealt{burgasser2010}), and cloud composition (e.g., \citealt{hiranaka2016}).  These properties can lead to objects having colors different than the ``normal'' population.  Color outliers include young sources, subdwarfs, binaries (both spectroscopic and visual), objects with uncertain spectral types, and objects with contaminated or suspect photometry.  We discuss each of these in more detail below.  All of the above samples are excluded from the list of objects used to determine typical UHS colors and absolute magnitudes of L, T, and Y dwarfs in Sections \ref{sec:colors} and Section \ref{sec:cmds}.  

\subsection{Binaries}
Binaries often stand out compared to the normal field sequence on color-spectral type and color-magnitude diagrams (CMDs).  L, T, and Y dwarf binaries in the UHS sample have generally been found in two ways, high angular-resolution imaging (e.g., \citealt{reid2001}) and spectral decomposition (e.g., \citealt{burgasser2010}).  We searched for evidence of resolved or spectral binarity in the literature for the entire UHS sample. Twenty-four objects were found to be resolved binaries \citep{reid2001, bouy2003, gizis2003, burgasser2005, burgasser2006b, reid2006, liu2006, siegler2007, burgasser2009, stumpf2010, artigau2011, liu2012, radigan2013}.  Note that we do not exclude wide binaries ($\gtrsim$1\arcsec) from color or absolute magnitude relations, as they are generally well-resolved in the photometric surveys used in this study.  For spectroscopic binaries, we only include those objects with strong evidence of binarity, and find twenty-four spectral binaries in the UHS sample \citep{burgasser2007, burgasser2010, geissler2011, kirkpatrick2011, mace2013, bardalez2014, best2015, robert2016, kellogg2017, bardalez2019, zhang2021}.  All resolved and strong spectroscopic binary candidates are flagged in Table \ref{tab:props}. 

A newly available indicator of binarity is the renormalized unit weight error (RUWE) from Gaia, which is a measure of the astrometric goodness of fit \citep{lindegren2018}.  RUWE values $>$1.4 are generally indicative of an object being a close binary \citep{lindegren2018}.  Twenty-two objects in our sample have a Gaia RUWE value $>$1.4, which are flagged in Table \ref{tab:props}.  This list includes the known, resolved binaries 2MASS J07003664+3157266AB (L3+L6.5, sep$\approx$0\farcs2; \citealt{reid2006, dupuy2012}), 2MASSW 0856479$+$223518 (L3+?, sep$\approx$0\farcs1; \citealt{bouy2003}), and Gl 417BC (L4.5+L6, sep$\approx$0\farcs07; \citealt{bouy2003, dupuy2012}). The sample of high-RUWE objects also includes spectroscopic binaries SDSS J080531.84$+$481233.0 (L4.5+T5; \citealt{burgasser2007}), SDSS J093113.23$+$280227.1 (L1.5+T2.5; \citealt{bardalez2014}), and 2MASS J11061197$+$2754225 (T0+T4.5; \citealt{burgasser2010}).  Note that SDSS J080531.84$+$481233.0 is also an astrometric binary \citep{dupuy2012, burgasser2016b, sahlmann2020}. Two known subdwarfs were also found to have high RUWE values, WISEA J043535.82$+$211508.9 (sdL0; \citealt{kirkpatrick2014}) and 2MASS J14343616$+$2202463 (d/sdL1; this work). It is possible that the exceptionally high proper motions of these objects compared to their parallaxes led to uncertain fits, or perhaps they are indeed subdwarf binaries.  Further investigation of these sources may be warranted. One additional wide companion was also found to have a RUWE value $>$1.4 (NLTT 44368B (L1.5); \citealt{deacon2014}).  Previous work has shown that wide companions are often found to be binaries themselves \citep{faherty2010, law2010}.  NLTT 44368B could be another example of this class of objects.  

\subsection{Red Photometric Outliers}
Young brown dwarfs (age $\lesssim$ 200 Myr) often have redder near-infrared colors than field-age counterparts of the same spectral type (e.g., \citealt{kirkpatrick2008}) due to their lower surface gravities.  However, not all redder-than-usual objects show clear signs of youth in their spectra (e.g., \citealt{looper2008, marocco2014}).  Eighty-one objects in the sample are flagged as young, red, or both in Table \ref{tab:props}.

\subsection{Blue Photometric Outliers}
Low metallicity objects (subdwarfs) often have unusually blue near-infrared colors compared to typical low-mass stars and brown dwarfs due to enhanced collision induced absorption (CIA) from H$_2$ (e.g., \citealt{linsky1969}). Though, not all low temperature objects with blue colors are subdwarfs (e.g., \citealt{cushing2010}).  Forty-four objects in this sample are flagged as being a subdwarf, blue, or both in Table \ref{tab:props}.

\subsection{Miscellaneous Outliers}
The spectra of some low-temperature objects are denoted as ``unusual'' or ``peculiar'' compared to spectral standards and do not easily fit into a category of known unusual objects (e.g., young, old, binary).  Such objects are often labeled as ``pec'' or given a ``:'' to indicate an uncertain spectral type determination, though the latter can also be applied to low-S/N sources.  All such objects are flagged in Table \ref{tab:props}.  Further, a small subset of objects in this sample have poor or contaminated photometry from one of the additional photometric surveys (e.g., Pan-STARRS DR2, CatWISE 2020) and are thus unsuitable for inclusion in determining photometric relations.  We also flag objects brighter than the nominal WFCAM saturation limit of $\sim$11 mag \citep{hodgkin2009}. Any objects with such photometry are flagged in Table \ref{tab:props} and are not included in any relations that use that particular photometry.

\subsection{Colors}
\label{sec:colors}

\begin{figure*}
\plotone{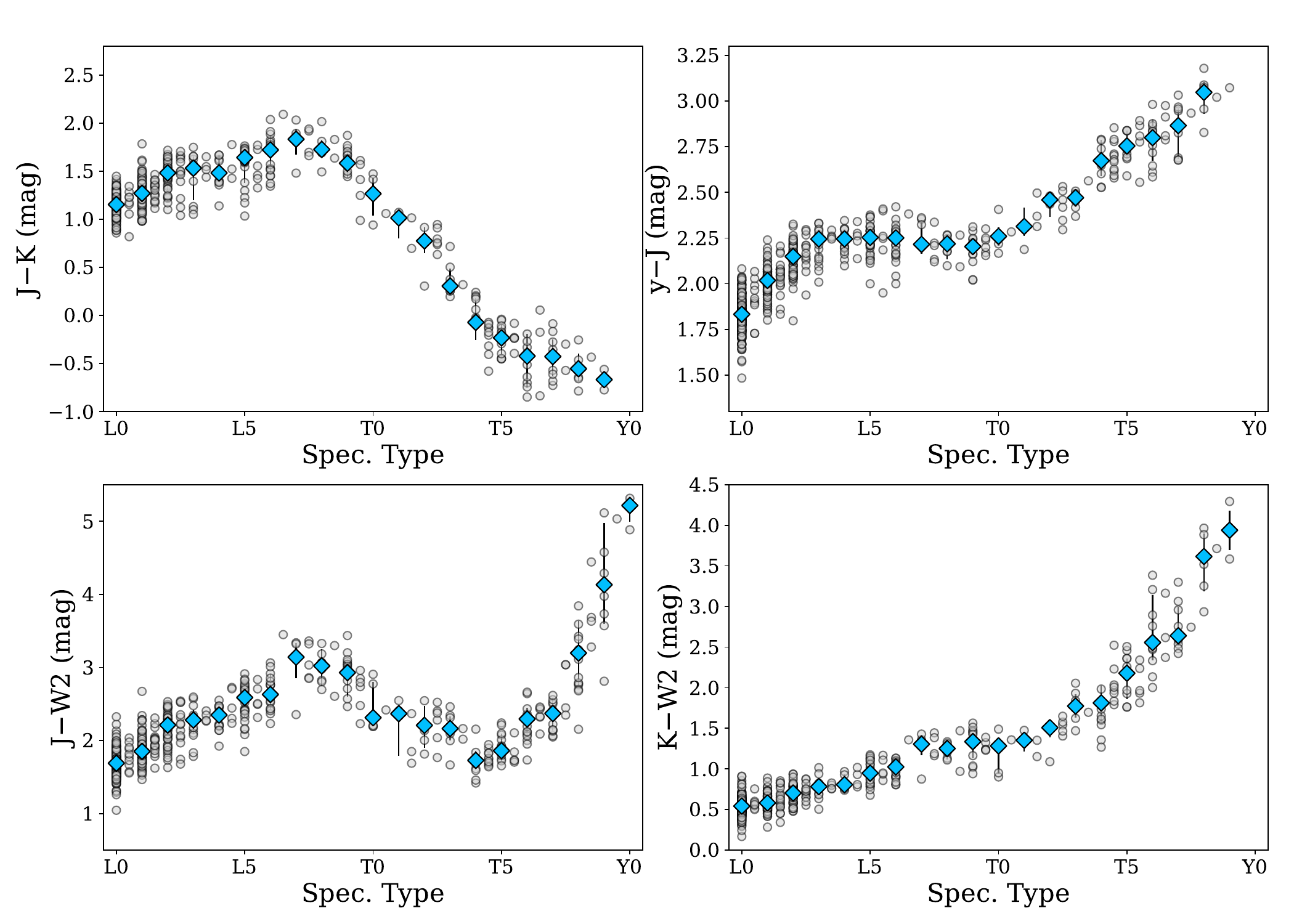}
\caption{UHS colors ($J-K$, $y-J$, $J-$W2, and $K-$W2) as a function of spectral type.  Median colors for integer types are shown as blue diamonds with errorbars showing the 16 and 84 percentile intervals and values presented in Table \ref{tab:colors}.}  
\label{fig:colors1}
\end{figure*}

Determining typical colors of L, T, and Y dwarfs is useful for characterizing such objects in general and as templates for estimated properties of newly discovered objects.  For this study, we determine the color versus spectral type trends for several commonly used colors involving UHS photometry: $J-K$, $y_{\rm PS1}-J$, $J-$W2 and $K-$W2, as shown in Figure \ref{fig:colors1}.  For colors that do not use UHS photometry, see previous studies for Pan-STARRS, WISE, or Gaia specific color information (e.g., \citealt{kirkpatrick2011, best2018, smart2019}). The median colors as a function of spectral type are provided in Table \ref{tab:colors}.  Note that we do not include objects flagged as young, red, subdwarf, blue, binary (spectroscopic or close visual), having uncertain or peculiar types, or poor/contaminated photometry.  The uncertainties listed in the table are the 16 and 84 percentile ranges and are only given when $\geq$5 objects are available for a particular spectral subtype.  When $<$ 5 objects are available for a particular subtype, we list only the median value.  

\begin{deluxetable*}{ccccccccc}
\label{tab:colors}
\tablecaption{Median Colors For L, T, and Y Dwarfs}
\tablehead{
\colhead{Spectral} & \colhead{N} & \colhead{$J-K$} & \colhead{N} & \colhead{$y-J$} & \colhead{N} & \colhead{$J-$W2} & \colhead{N} & \colhead{$K-$W2} \\
\colhead{Type} & \colhead{} & \colhead{(mag)} & \colhead{} & \colhead{(mag)} & \colhead{} & \colhead{(mag)} & \colhead{} & \colhead{(mag)} }
\startdata
L0--L0.5 & 218 & 1.15$^{+0.10}_{-0.10}$ & 263 & 1.83$^{+0.10}_{-0.09}$ & 263 & 1.69$^{+0.15}_{-0.14}$ & 219 & 0.54$^{+0.09}_{-0.09}$ \\
L1--L1.5 & 112 & 1.27$^{+0.14}_{-0.13}$ & 127 & 2.02$^{+0.06}_{-0.08}$ & 125 & 1.85$^{+0.18}_{-0.14}$ & 113 & 0.58$^{+0.10}_{-0.08}$ \\
L2--L2.5 & 56 & 1.48$^{+0.14}_{-0.18}$ & 67 & 2.15$^{+0.08}_{-0.10}$ & 67 & 2.21$^{+0.18}_{-0.30}$ & 57 & 0.70$^{+0.11}_{-0.13}$ \\
L3--L3.5 & 15 & 1.53$^{+0.12}_{-0.34}$ & 19 & 2.24$^{+0.05}_{-0.13}$ & 19 & 2.28$^{+0.13}_{-0.18}$ & 15 & 0.78$^{+0.04}_{-0.10}$ \\
L4--L4.5 & 14 & 1.48$^{+0.18}_{-0.10}$ & 18 & 2.25$^{+0.05}_{-0.09}$ & 18 & 2.35$^{+0.15}_{-0.17}$ & 15 & 0.81$^{+0.12}_{-0.03}$ \\
L5--L5.5 & 32 & 1.64$^{+0.07}_{-0.26}$ & 36 & 2.25$^{+0.07}_{-0.10}$ & 35 & 2.59$^{+0.16}_{-0.27}$ & 30 & 0.95$^{+0.17}_{-0.13}$ \\
L6--L6.5 & 21 & 1.72$^{+0.15}_{-0.26}$ & 22 & 2.25$^{+0.06}_{-0.10}$ & 21 & 2.63$^{+0.27}_{-0.20}$ & 20 & 1.02$^{+0.07}_{-0.14}$ \\
L7--L7.5 & 9 & 1.83$^{+0.10}_{-0.16}$ & 10 & 2.22$^{+0.13}_{-0.05}$ & 10 & 3.14$^{+0.19}_{-0.29}$ & 9 & 1.30$^{+0.11}_{-0.14}$ \\
L8--L8.5 & 9 & 1.73$^{+0.10}_{-0.07}$ & 10 & 2.22$^{+0.05}_{-0.09}$ & 10 & 3.02$^{+0.23}_{-0.28}$ & 9 & 1.25$^{+0.08}_{-0.14}$ \\
L9--L9.5 & 18 & 1.58$^{+0.13}_{-0.15}$ & 21 & 2.20$^{+0.05}_{-0.05}$ & 21 & 2.93$^{+0.14}_{-0.33}$ & 18 & 1.34$^{+0.13}_{-0.21}$ \\
T0--T0.5 & 6 & 1.27$^{+0.16}_{-0.23}$ & 6 & 2.26$^{+0.05}_{-0.05}$ & 6 & 2.31$^{+0.49}_{-0.12}$ & 6 & 1.28$^{+0.10}_{-0.34}$ \\
T1--T1.5 & 3 & 1.01 & 5 & 2.31$^{+0.10}_{-0.05}$ & 5 & 2.37$^{+0.09}_{-0.58}$ & 3 & 1.35 \\
T2--T2.5 & 8 & 0.77$^{+0.14}_{-0.13}$ & 9 & 2.46$^{+0.04}_{-0.09}$ & 10 & 2.21$^{+0.26}_{-0.31}$ & 7 & 1.51$^{+0.07}_{-0.12}$ \\
T3--T3.5 & 8 & 0.31$^{+0.18}_{-0.05}$ & 8 & 2.47$^{+0.03}_{-0.05}$ & 9 & 2.16$^{+0.17}_{-0.16}$ & 8 & 1.77$^{+0.15}_{-0.14}$ \\
T4--T4.5 & 17 & -0.07$^{+0.25}_{-0.18}$ & 18 & 2.67$^{+0.11}_{-0.08}$ & 18 & 1.73$^{+0.13}_{-0.12}$ & 16 & 1.81$^{+0.21}_{-0.25}$ \\
T5--T5.5 & 16 & -0.23$^{+0.14}_{-0.20}$ & 17 & 2.75$^{+0.08}_{-0.07}$ & 19 & 1.86$^{+0.24}_{-0.14}$ & 16 & 2.18$^{+0.18}_{-0.32}$ \\
T6--T6.5 & 13 & -0.42$^{+0.23}_{-0.33}$ & 16 & 2.80$^{+0.10}_{-0.13}$ & 19 & 2.29$^{+0.15}_{-0.24}$ & 14 & 2.56$^{+0.59}_{-0.22}$ \\
T7--T7.5 & 12 & -0.43$^{+0.18}_{-0.20}$ & 11 & 2.87$^{+0.10}_{-0.18}$ & 17 & 2.37$^{+0.21}_{-0.23}$ & 11 & 2.64$^{+0.36}_{-0.12}$ \\
T8--T8.5 & 6 & -0.55$^{+0.16}_{-0.13}$ & 6 & 3.05$^{+0.06}_{-0.12}$ & 19 & 3.20$^{+0.44}_{-0.44}$ & 6 & 3.62$^{+0.28}_{-0.43}$ \\
T9--T9.5 & 2 & -0.67 & 1 & 3.07 & 8 & 4.13$^{+0.85}_{-0.54}$ & 2 & 3.94 \\
Y0 & 0 & \dots & 0 & \dots & 3 & 5.22 & 0 & \dots \\
\enddata
\tablecomments{This table gives the number of objects followed by median colors or L, T, and Y dwarfs in UHS.  Uncertainties are the 16 and 84 percentile ranges, and are only calculated when N $\geq$ 5. }
\end{deluxetable*}

\subsection{CMDs}
\label{sec:cmds}

Color-magnitude diagrams are also useful for characterizing low-temperature samples. They are also a tool for finding relationships that can be extended to objects without measured parallaxes in order to determine photometric distances.  We have gathered measured parallaxes for every object in our sample, which are provided in Table \ref{tab:props} \citep{dahn2002, tinney2003, vrba2004, schilbach2009, dupuy2012, manjavacas2013, smart2013, zapatero2014, dupuy2015, liu2016, sahlmann2016, dahn2017, dupuy2017, gaia2018, kirkpatrick2019, lodieu2019, best2020, kirkpatrick2021, zhang2021, gaia2022}.  If more than one parallax measurement is available for a particular object, we use whichever has the smallest uncertainty.  

\begin{figure*}
\plotone{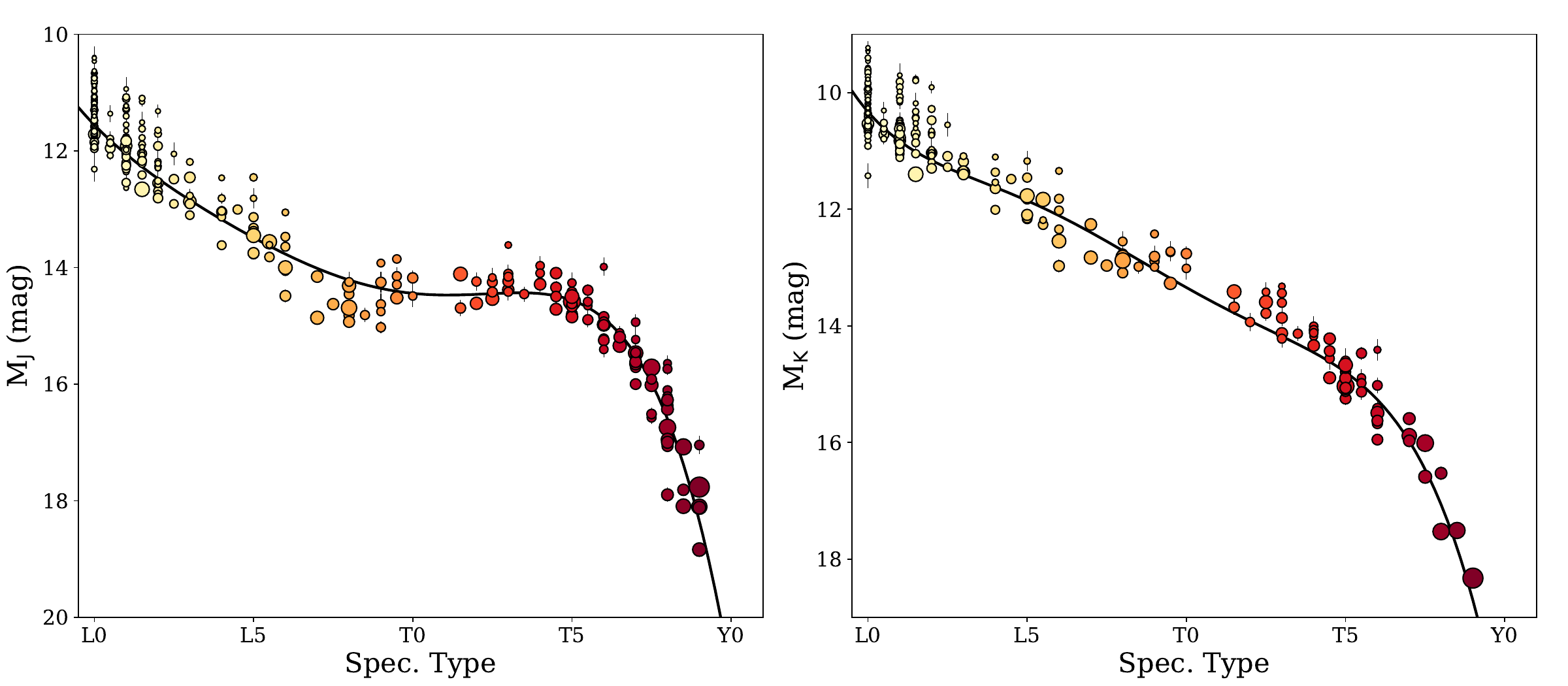}
\caption{Absolute UHS $J$- and $K$-band magnitudes as a function of spectral type for L, T, and Y dwarfs observed as part of the UKIRT Hemisphere Survey.  Symbol sizes represent distances, with larger symbols corresponding to closer distances.  Colors correspond to spectral type.  This figure does not include known close binaries (spectroscopic or resolved), young objects, subdwarfs, objects with uncertain types, objects designated ``red'', ``blue'', or ``pec'', or objects with a parallax S/N $<$ 10. }  
\label{fig:absmags}
\end{figure*}

Absolute $J$- and $K$-band magnitudes versus spectral type trends are shown in Figure \ref{fig:absmags}.  As with the color-spectral type trends, we do not include objects flagged as young, red, subdwarf, blue, binary (spectroscopic or close visual), having uncertain or peculiar types, or poor/contaminated photometry in Table \ref{tab:props}.  In addition, we require parallax S/N $>$10 and a photometric S/N $>$ 5.  We fit the trends for $J$ and $K$ with a weighted fifth order polynomial and give the coefficients of the fits in Table \ref{tab:coeffs}.  

\begin{deluxetable*}{ccccccccc}
\label{tab:coeffs}
\tablecaption{Coefficients of Absolute Magnitude Polynomial Fits}
\tablehead{
\colhead{x} & \colhead{y} & \colhead{$c_0$} & \colhead{$c_1$} & \colhead{$c_2$} & \colhead{$c_3$} & \colhead{$c_4$} & \colhead{$c_5$} & \colhead{rms} \\
}
\startdata
SpT & $M_{\rm J}$ & -2.35283E+01 & 1.01992E+01 & -1.21981E+00 & 7.54721E-02 & -2.31859E-03 & 2.78465E-05 & 0.406 \\
SpT & $M_{\rm K}$ & -5.36375E+01 & 1.87805E+01 & -2.17131E+00 & 1.24194E-01 & -3.47013E-03 & 3.79874E-05 & 0.399 \\
\enddata
\tablecomments{These polynomials take the form \[y = \sum_{i=0}^{n} c_i x^i ,\] where spectral type L0 = 10, T0 = 20, and Y0 = 30.}
\end{deluxetable*}

\section{Astrometry}
\label{sec:pms}

To measure proper motions for known L, T, and Y dwarfs in the UHS survey, we required both a $J$- and $K$-band detection in the UHS DR2 catalog, which ensures a sufficient time baseline for a proper motion measurement.  Out of the full 966-object sample, 14 sources have $K$-band detections but no corresponding $J$-band detection.  While the $J$-band UHS survey area is generally more complete, there are some small regions where there is $K$-band coverage and no $J$-band coverage. All 14 $K$-only sources are due to these small coverage differences. There are a total of 172 objects that have $J$-band detections, but no corresponding $K$-band detection.  Some of these are due to the more complete coverage of the $J$-band survey compared to $K$, which should be filled in with future UHS data releases. There is a significant fraction of the $J$-only sample, however, that does have $K$-band coverage but remains undetected.  These objects typically have spectral types $\gtrsim$T5, where $J-K$ colors turn especially blue.   

The formal data releases of UHS are astrometrically calibrated against 2MASS.  Bright sources in 2MASS have an astrometric accuracy of around 100 mas and lack proper motions \citep{skrutskie2006}.  In order to improve the astrometry, we have recalibrated each pawprint (a pawprint is the set of four images, one for each detector in the camera, for a single exposure) from the wide-field camera (WFCAM; \citealt{casali2007}) against Gaia DR3, which has better than 1 mas astrometric accuracy, including proper motions and parallaxes, to G$\sim$20 \cite{gaia2022}.  Further, distortions in the WFCAM focal plane were removed by generating residual maps and applying those corrections.  Separate residual maps were used for UHS-$J$ and UHS-$K$, but each map was found to be stable throughout the survey.   The rms residual between the recalibrated coordinates and Gaia for the bright stars used for the recalibrations is typically around 8 mas, with most better than 15 mas.

The recalibrated $J$- and $K$- band positions are then used to derive proper motions.  For the objects with both $J$- and $K$-band detections (782 objects total), the median time baseline is $\sim$5.1 years between observations.  If a matching Pan-STARRS detection is found, its position is used as well in deriving the proper motions. Each position is weighted by its inverse variance in the proper motion calculation.  Gaia positions are not used, as if a given object has a Gaia match, then the Gaia proper motion is preferred over the proper motions calculated here.  All proper motions derived in this way are provided in Table \ref{tab:props}.

\begin{figure*}
\plotone{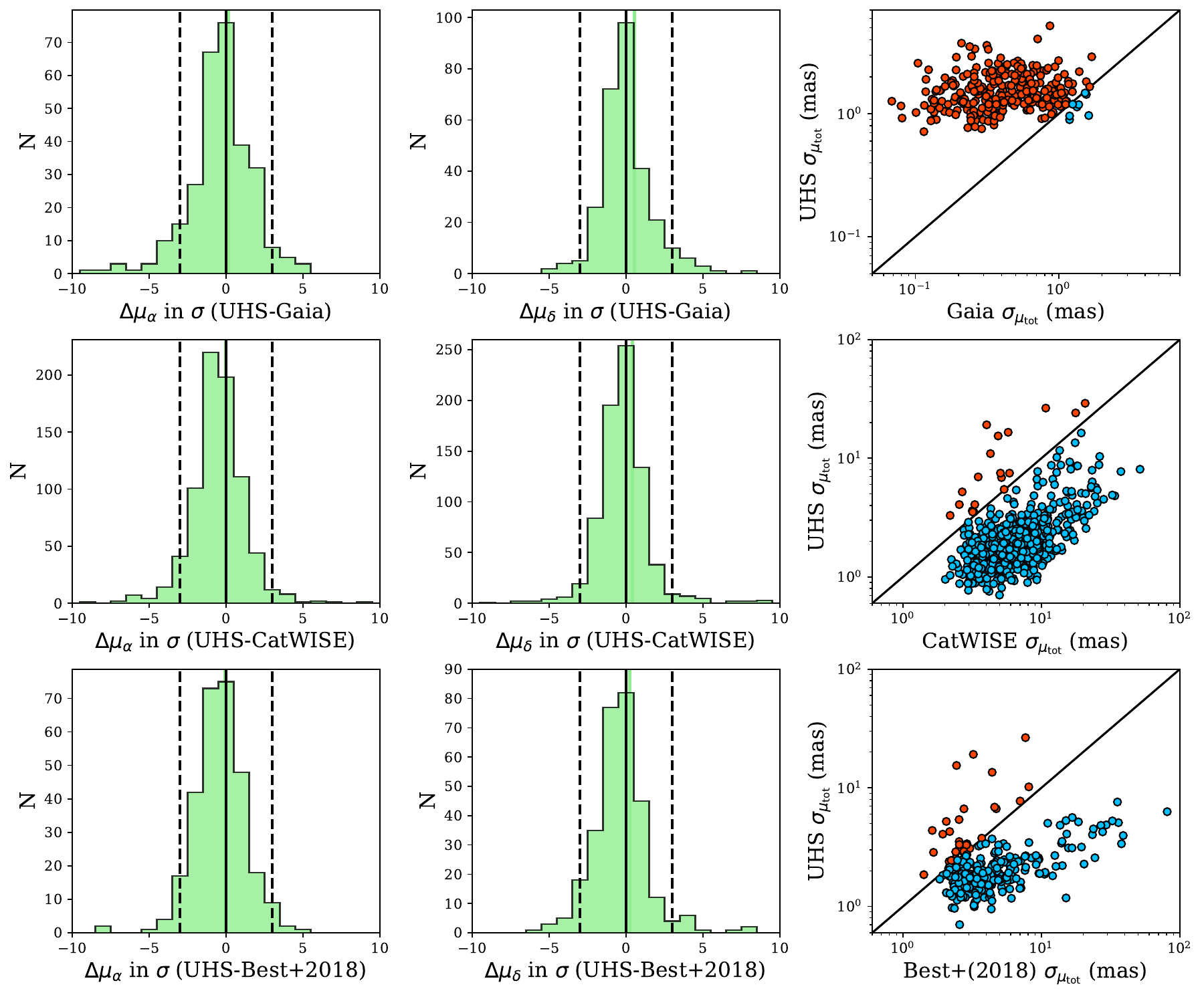}
\caption{A comparison of proper motions derived from UHS positions in this work to Gaia DR3 (top row), CatWISE 2020 (middle row) and Pan-STARRS derived proper motions in \cite{best2018} (bottom row).  The left panels show histograms of the difference between $\mu_{\alpha}$ measurements in combined $\sigma$, with dashed black lines indicating 3$\sigma$ differences between measurements.  The middle panels show similar histograms for $\mu_{\delta}$.  The right panels compare the uncertainty values of the total proper motions for objects, with red symbols indicating smaller uncertainties in either Gaia DR3, CatWISE 2020, or \cite{best2018} compared to UHS proper motions, and blue symbols indicating smaller uncertainties for UHS proper motions.  These are separated by a solid diagonal line indicating 1-to-1 values. }  
\label{fig:pms1}
\end{figure*}

In order to validate and evaluate the performance of the proper motions determined using UHS data, we compare our measurements to three different sources: Gaia DR3 \citep{gaia2022}, CatWISE 2020 \citep{marocco2021}, and the Pan-STARRS derived proper motions in \cite{best2018}.  The main results of this comparison are shown in Figure \ref{fig:pms1}.

We find 291 objects from our proper motion sample that have proper motions in Gaia DR3 \citep{gaia2022}, as shown in the top panels of Figure \ref{fig:pms1}.  The vast majority of these sources have higher precision measurements in Gaia compared to our UHS proper motions (283/291; 97.2\%).  We find generally good agreement, with 80.8\% of objects having both proper motion components within 3$\sigma$ of Gaia DR3 measurements.  

We find 774 objects from our proper motion sample with proper motions in CatWISE 2020 \citep{marocco2021}, shown in the middle row of Figure \ref{fig:pms1}.  Of these, the UHS derived proper motions have a higher precision for 755 objects (97.5\%).  Again, we generally find good agreement, with 88.4\% of objects in common having both proper motion components within 3$\sigma$.

Because Pan-STARRS DR2 proper motions are not publicly available, we compare to the Pan-STARRS derived motions determined in \cite{best2018}, which can be seen in the bottom row of Figure \ref{fig:pms1}.  There are 293 objects with proper motions in \cite{best2018} in common with our UHS proper motion sample.  We find higher precision from the UHS derived proper motions for 262 (89.4\%) of these objects.  There is good agreement, with 87.4\% of objects in common having both proper motion components within 3$\sigma$.  

We have further searched the literature and available catalogs to find the most precise proper motions that exist for all of the objects in the full UHS LTY sample.  Any proper motion found that is more precise than available proper motions from this work, Gaia DR3, CatWISE 2020, and \cite{best2018} is provided in Table \ref{tab:props}.  We find that our UHS derived proper motions are the most precise measurements for 381 of the 966 objects in this work ($\sim$39\%).  For the investigations in Section \ref{sec:ooi}, we use the best available proper motions.

\section{Objects of Interest}
\label{sec:ooi}

\subsection{Young Moving Group Members}
\label{sec:mgs}

Nearby associations and moving groups in the solar neighborhood serve as important laboratories for testing stellar and substellar evolutionary theory.  Any L, T, or Y dwarf that can be tied to such a group with a well-determined age becomes an important benchmark, as ages are typically difficult to determine for field L, T, and Y dwarfs.  Nearby groups tend to have distinct kinematics, and several tools have been developed to determine probabilities of belonging to various nearby groups based on kinematics alone.

The best available astrometry for all objects in this sample was input into BANYAN $\Sigma$ \citep{gagne2018} and LACEwING \citep{riedel2017} to determine potential moving group or association membership.  BANYAN $\Sigma$ evaluates potential membership for 29 groups in the solar neighborhood via a Bayesian classifier and a candidate's available astrometric information (position, proper motion, parallax, and radial velocity).  Alternatively, LACEwING uses a frequentist framework to compare the kinematic information of a candidate to 16 nearby moving groups.  All of the groups included in LACEwING are also included in BANYAN $\Sigma$, though the lists of known members varies.  For both BANYAN $\Sigma$ and LACEwING, we choose a relatively conservative probability threshold value of $\geq$80\% for an object to be considered a candidate moving group or association member.  For each object, we used UHS positions, the best available proper motion values (based on smallest uncertainties), and the highest-precision parallax (when available).  If a parallax is not available, we use $J$-band photometric distances and corresponding uncertainties determined from Table \ref{tab:coeffs}.  $K$-band photometric distances are used if measured parallaxes and $J$-band photometry are unavailable.   

Eighty-seven objects were found to have membership probabilities of 80\% or larger from BANYAN $\Sigma$, LACEwING, or both.  Fifty-four of these are either known members or previously suggested candidate members.  We discuss each group for which a member or candidate member was found, and evaluate potential membership for new candidates in the following sections.

\subsubsection{The AB Doradus Moving Group}

The AB Doradus Moving Group (ABDMG) was first identified in \cite{zuckerman2004} and has an estimated age of 149$^{+51}_{-19}$ Myr \citep{bell2015}.  Our search recovered nine known or previously suggested candidate ABDMG members and six new candidate members, listed in Table \ref{tab:abdmg}.  Two recovered candidates are the wide companions LP 261-75B and HD 89744B.  \cite{liu2016} found the L 261-75AB system to most likely belong to the young field population and we adopt that result here, listing L 261-75B in the rejected members in Table \ref{tab:abdmg}.  Further, the primary to HD 89744B (HD 89744A) was once considered an AB Dor member, but later found to have an age of several Gyr, inconsistent with AB Dor membership \citep{schaefer2018}.  We adopt this result here and list HD 89744B in the list of rejected members.  Recovered members include the very well-known low-mass members 2MASS J03552337$+$1133437 and 2MASSW J2244316$+$204343, as well as the planetary mass companion GU Psc b \citep{naud2014}.  Our UHS proper motion of GU Psc b is the most precise measured so far for this object.  Using \texttt{COMOVER} \citep{gagne2021} and Gaia DR3 astrometry for the primary, we find a co-moving probability of 98.6\% of GU Psc and GU Psc b without using a distance estimate for the b component.  If we include the $K$-band distance estimate of 44$\pm$8 pc (determined using the relations given in Table \ref{tab:coeffs}), we find a comoving probability of 99.98\%.  While the physical association of GU Psc and GU Psc b is not in doubt, our measured proper motion for the b component is consistent with and reaffirms the physical association of this pair.  

Of the six new candidate members identified in this work, we reject three of them as possible ABDMG members.  CWISE J043309.36$+$100902.3 returns a significant probability of ABDMG membership from both BANYAN and LACEwING, however, the best proper motion values for this object, which come from \cite{meisner2020}, have relatively large uncertainties ($\sim$35 mas yr$^{-1}$).  A more precise proper motion is needed for this object to consider it an ABDMG candidate member. PSO J004.1834$+$23.0741 was returned as a high-probability ABDMG member from LACEwING, but has a 0\% probability of ABDMG membership from BANYAN $\Sigma$, and we therefore do not consider this object as a new candidate ABDMG member. WISE J111838.70$+$312537.9 is a widely separated companion to the quadruple star system $\xi$ Ursae Majoris, which was found to have an age of several Gyr \citep{wright2013}, and is thus not an ABDMG member   

Both 2MASSI J0409095$+$210439 and 2MASS J06143818$+$3950357 are returned as high-probability ABDMG members according to BANYAN $\Sigma$, with non-zero probabilities of ABDMG from LACEwING as well.  Both objects have well-measured parallaxes, which bolsters their potential membership.   We retrieved the IRTF/SpeX spectrum of 2MASSI J0409095$+$210439, and while it has no clear spectral features indicating a low surface gravity, we note that it is slightly red compared to the L3 and L4 standards, a common feature of known young brown dwarfs.  The near-infrared spectrum of 2MASS J06143818$+$3950357 presented in \cite{muzic2012} is a good match to the L9 standard with no obvious signs of low-gravity.  LACEwING also gives a high probability of belonging to the Hyades for this object (88.0\%), but we find membership unlikely in Section \ref{sec:hya}.  Higher-resolution spectroscopy of these objects would allow for a deeper investigation of gravity sensitive features and radial velocity measurements, which would firmly assess membership status for these L dwarfs.  Both objects are flagged as potentially young in Table \ref{tab:props}.

PSO J057.2893$+$15.2433 is a red L7 discovered in \cite{best2015}, who found a strong probability of membership in the $\beta$ Pictoris Moving Group (BPMG).  Using our measured UHS proper motion and a $J$-band photometric distance, we instead find a high probability of ABDMG membership.  We note, however, that $J$-band photometric distances tend to underestimate actual distances for red L dwarfs because of their unusual SEDs \citep{schneider2023}.  Using the $K$-band photometric distance for this object (31$\pm$5 pc, compared to 48$\pm$8 pc for $J$) we find ambiguous membership from BANYAN $\Sigma$ -- 54.1\% membership probability in ABDMG and 37.9\% membership probability in BPMG.  We retain this object as an ABDMG candidate member in need of parallax and radial velocity measurements.

\begin{longrotatetable}
\begin{deluxetable*}{lccccccccccccccc}
\label{tab:abdmg}
\tablecaption{AB Dor Members and Candidate Members }
\tablehead{
\colhead{Name} & \colhead{Sp.} & \colhead{Ref.} & \colhead{$\mu_{\alpha}$} & \colhead{$\mu_{\delta}$} & \colhead{Ref.} & \colhead{$\varpi$\tablenotemark{a}} & \colhead{Ref.} & \colhead{BANYAN} & \colhead{LACEwING} &\colhead{Ref.\tablenotemark{b}}  \\
\colhead{          } & \colhead{Type} & \colhead{      } & \colhead{(mas yr$^{-1}$)} & \colhead{(mas yr$^{-1}$)} & \colhead{ } & \colhead{(mas)} & \colhead{ } & \colhead{(\%)} & \colhead{(\%)} &\colhead{ }  }
\startdata
\cutinhead{Recovered Members}
GU Psc b & T3.5 & 1 & 77.3$\pm$13.2 & -101.0$\pm$13.1 & 2 & 21.01$\pm$0.03\tablenotemark{c} & 3 & 92.0 & 13.0 & 1,4,5 \\
2MASS J03552337$+$1133437 & L3--L6$\gamma$ & 6 & 223.18$\pm$0.59 & -631.30$\pm$0.37 & 3 & 109.14$\pm$0.48 & 3 & 99.7 & 87.0 & 7,8,9,10,11,12 \\
2MASS J04203904$+$2355502 & L1 & 13 & 44.72$\pm$0.69 & -168.61$\pm$0.46 & 3 & 25.17$\pm$0.60 & 3 & 98.3 & 50.0 & 14,15 \\
WISE J064205.58$+$410155.5 & L9 (red) & 16 & -2.0$\pm$1.2 & -383.1$\pm$1.2 & 17 & 62.6$\pm$3.1 & 17 & 85.9 & 66.0 & 17 \\
PSO J318.4243$+$35.1277 & L1 & 4 & 111.73$\pm$0.25 & -76.86$\pm$0.27 & 3 & 32.71$\pm$0.30 & 3 & 98.7 & 30.0 & 4,18 \\
2MASSW J2244316$+$204343 & L6-L8$\gamma$ & 6 & 230.3$\pm$0.9 & -234.8$\pm$1.0 & 12 & 58.7$\pm$1.0 & 12 & 99.7 & 54.0 & 4,10,11,12,19 \\
PSO J358.5527$+$22.1393 & L2 & 4 & 101.22$\pm$0.67 & -89.30$\pm$0.40 & 3 & 22.98$\pm$0.61 & 3 & 93.4 & 21.0 & 4,18 \\
\cutinhead{New Candidate Members}
PSO J057.2893$+$15.2433 & L7 (red) & 16 & 72.33$\pm$3.83 & -133.91$\pm$3.58 & 2 & [21$\pm$4] & 2 & 89.8 & 59.0 & 16 \\
2MASSI J0409095$+$210439 & L3.5 & 20 & 92.51$\pm$1.07 & -171.28$\pm$0.81 & 3 & 31.49$\pm$0.92 & 3 & 89.9 & 74.0 & \dots \\ 
2MASS J06143818$+$3950357 & L9 & 21 & -31.9$\pm$1.7 & -264.9$\pm$1.8 & 22 & 44.0$\pm$2.6 & 22 & 85.8 & 58.0 & \dots \\
\cutinhead{Rejected Candidate Members}
PSO J004.1834$+$23.0741 & T0 & 16 & 405.4$\pm$2.1 & 61.8$\pm$2.5 & 22 & 38.4$\pm$3.3 & 22 & 0.0 & 86.0 & \dots \\
CWISE J043309.36$+$100902.3 & T8 & 17 & 174$\pm$34 & -384$\pm$36 & 23 & [53$\pm$9] & 2 & 88.7 & 68.0 & \dots \\
LP 261-75B & L6: & 12 & -94.0$\pm$2.4 & -164.3$\pm$2.7 & 12 & 29.6$\pm$2.8 & 12 & 83.7 & 25.0 & 12 \\
HD 89744B & L0.5 & 24 & -119.22$\pm$0.65 & -140.39$\pm$0.46 & 3 & 26.01$\pm$47 & 3 & 91.6 & 13.0 & 25 \\
WISE J111838.70$+$312537.9 & T8.5 & 26 & -405.1$\pm$8.1 & -588.6$\pm$9.4 & 27 & 114.49$\pm$0.43\tablenotemark{d} & 3 & 84.0 & 39.0 & \dots \\
\enddata
\tablenotetext{a}{Parallaxes in square brackets are photometric. }
\tablenotetext{b}{References in the final column are the relevant works that have previously evaluated, suggested, or determined moving group membership. }
\tablenotetext{c}{The parallax for this object comes from the primary star GU Psc. }
\tablenotetext{d}{The parallax for this object comes from the primary star $\xi$ Uma. }
\tablerefs{(1) \cite{naud2014}; (2) This work; (3) \cite{gaia2022}; (4) \cite{aller2016}; (5) \cite{zhang2021}; (6) \cite{gagne2015}; (7) \cite{faherty2013}; (8) \cite{liu2013}; (9) \cite{zapatero2014}; (10) \cite{gagne2014}; (11) \cite{faherty2016}; (12) \cite{liu2016}; (13) \cite{luhman2006}; (14) \cite{kraus2017}; (15) \cite{gagne2018};  (16) \cite{best2015}; (17) \cite{kirkpatrick2021}; (18) \cite{gagne2018b}; (19) \cite{vos2018}; (20) \cite{bardalez2014}; (21) \cite{muzic2012}; (22) \cite{best2020}; (23) \cite{meisner2020}; (24) \cite{schneider2014}; (25) \cite{schaefer2018}; (26) \cite{wright2013}; (27) \citep{marocco2021}} 
\end{deluxetable*}
\end{longrotatetable}

\subsubsection{The Argus Association}
\label{sec:arg}

The Argus Association (ARG) was originally identified in \cite{torres2008}, though this original group was heavily contaminated, leading \cite{zuckerman2019} to reexamine and redefine the membership of this group and give an age of 40--50 Myr.  Our search recovered 2 known L-type members, and returned 7 new candidate members, summarized in Table \ref{tab:argus}. 

The new ARG candidate 2MASS J04070752$+$1546457 is a strong H$\alpha$ emitter \citep{reid2008} and a rapid rotator \citep{tannock2021}.  Despite being a high-probability ARG member from BANYAN $\Sigma$ and a moderate-probability member according to LACEwING, the predicted radial velocities ($\sim$21.6 \kms\ and $\sim$21.1 \kms, respectively) do not agree with the measured radial velocity from \cite{tannock2021} (43.4$\pm$2.1 \kms).  We therefore reject 2MASS J04070752$+$1546457 as an ARG member.

The new candidate CWISE J062317.13$+$263129.7 was discovered by \cite{scholz2020} using Gaia data and given a photometric type of L4.5$\pm$2.5.  This object was spectroscopically classified in \cite{kirkpatrick2021}, and labeled as ``L3 pec (composite?)''.  As seen in Figure 10 of that paper, the spectrum is a decent match to the L3 standard at $J$, but much redder at $H$ and $K$, possibly indicating a low surface gravity.  This object was not evaluated as a potential moving group member in \cite{kirkpatrick2021} because that investigation was limited to objects determined to be within 20 pc.  A radial velocity is needed to establish ARG membership for this object.

BANYAN $\Sigma$ returned high membership ARG probabilities for 2MASS J15311344$+$1641282, 2MASS J17153111$+$1054108,  SDSS J202820.32$+$005226.5, 2MASS J23313131$+$2041273.  The IRTF/SpeX spectra of these sources show no obvious signs of low surface gravity, but there is no clear reason to rule these objects out as candidates. More information is needed in order to establish ARG or field membership for these objects.  We retain all four objects as ARG candidates.

SDSS J135923.99$+$472843.2 was discovered in \cite{knapp2004} and assigned a near-infrared type of L8.5.  There is no previous evidence of youth for this object in the literature, but it has not been studied in great detail.  Our UHS proper motion suggests ARG membership and we keep it as a candidate.

We consider CWISE J062317.13$+$263129.7, SDSS J135923.99$+$472843.2, 2MASS J15311344$+$1641282, 2MASS J17153111$+$1054108, SDSS J202820.32$+$005226.5, and 2MASS J23313131+2041273 potential new ARG members, which are flagged as possibly young in Table \ref{tab:props}.

Finally, we mention 2MASS J10271549$+$5445175, which was determined to have a spectral type of L2 VL-G in Section \ref{sec:sample}.  BANYAN $\Sigma$ returns a  46.5\% probability for ARG membership, with no other groups having probabilities $>$1\%.  The predicted distance for ARG if a member is $\sim$53.4 pc, which is in good agreement with the photometric $J$-band distance for this object of 50$\pm$9 pc (note that no UHS $K$-band photometry is available for this source).  Because the spectrum shows signs of low gravity and the estimated distance for this object aligns well with ARG membership, we consider this object an ARG candidate member in need of a more detailed investigation.

\begin{longrotatetable}
\begin{deluxetable*}{lccccccccccccccc}
\label{tab:argus}
\tablecaption{Argus Members and Candidate Members }
\tablehead{
\colhead{Name} & \colhead{Sp.} & \colhead{Ref.} & \colhead{$\mu_{\alpha}$} & \colhead{$\mu_{\delta}$} & \colhead{Ref.} & \colhead{$\varpi$\tablenotemark{a}} & \colhead{Ref.} & \colhead{BANYAN} & \colhead{LACEwING} &\colhead{Ref.\tablenotemark{b}}  \\
\colhead{          } & \colhead{Type} & \colhead{      } & \colhead{(mas yr$^{-1}$)} & \colhead{(mas yr$^{-1}$)} & \colhead{ } & \colhead{(mas)} & \colhead{ } & \colhead{(\%)} & \colhead{(\%)} &\colhead{ }  }
\startdata
\cutinhead{Recovered Members}
2MASSW J0045214$+$163445 & L2 VL-G & 1 & 359.07$\pm$0.20 & -47.91$\pm$0.14 & 2 & 65.406$\pm$0.175 & 2 & 99.1 & 67.0 & 3,4,5,6,7,8,9,10 \\
SDSS J213240.36$+$102949.4 & L4: $\beta$ & 11 & 107.94$\pm$2.23 & 27.67$\pm$2.30 & 12 & [23$\pm$4] & 12 & 84.2 & 5.0 & 6,11,13 \\
\cutinhead{New Candidate Members}
CWISE J062317.13$+$263129.7 & L3 pec & 10 & -8.56$\pm$0.54 & -130.33$\pm$0.39 & 2 & 48.61$\pm$0.48 & 2 & 99.1 & 52.0 & \dots \\ 
SDSS J135923.99$+$472843.2 & L8.5 & 14 & -158.27$\pm$2.36 & 59.05$\pm$2.51 & 12 & [28$\pm$5] & 12 & 85.1 & 0.0 & \dots \\
2MASS J15311344$+$1641282 & L2 & 12 & -89.74$\pm$1.43 & 29.70$\pm$0.99 & 2 & 26.75$\pm$1.12 & 2 & 96.2 & 0.0 & \dots \\
2MASS J17153111$+$1054108 & L6 & 15 & -38.59$\pm$2.93 & 2.68$\pm$2.76 & 12 & [25$\pm$4] & 12 & 88.7 & 6.0 & \dots \\
SDSS J202820.32$+$005226.5 & L2 & 16 & 96.04$\pm$0.54 & -11.16$\pm$0.44 & 2 & 35.14$\pm$0.45 & 17 & 98.3 & 44.0 & \dots \\
2MASS J23313131$+$2041273 & L3.5 & 2 & 124.74$\pm$2.16 & -10.79$\pm$2.22 & 12 & [26$\pm$5] & 12 & 89.6 & 0.0 & \dots \\
\cutinhead{Rejected Candidate Members}
2MASS J04070752$+$1546457 & L4 (red) & 12 & 74.9$\pm$1.5 & -64.3$\pm$1.1 & 7 & 28.9$\pm$1.3 & 10 & 84.6 & 32.0 & \dots \\
\enddata
\tablenotetext{a}{Parallaxes in square brackets are photometric. }
\tablenotetext{b}{References in the final column are the relevant works that have previously evaluated, suggested, or determined moving group membership. }
\tablerefs{(1) \cite{allers2013}; (2) \cite{gaia2022}; (3) \cite{gagne2014}; (4) \cite{zapatero2014}; (5) \cite{gagne2015b}; (6) \cite{faherty2016}; (7) \cite{liu2016}; (8) \cite{riedel2019}; (9) \cite{ujjwal2020}; (10) \cite{kirkpatrick2021}; (11) \cite{gagne2015}; (12) This work; (13) \cite{vos2019}; (14) \cite{knapp2004}; (15) \cite{kellogg2017}; (16) \cite{burgasser2010}; (17) \cite{dahn2017}} 
\end{deluxetable*}
\end{longrotatetable}

\subsubsection{The $\beta$ Pictoris Moving Group}
The Beta Pictoris Moving Group (BPMG) was identified in \cite{zuckerman2001}, with the most recent age determination of 20.4$\pm$2.5 Myr \citep{couture2023}.  Our search returned one previously known candidate BPMG member (CWISE J050626.96$+$073842.4; \citealt{schneider2023}).  No new BPMG candidates were found in our search.

\subsubsection{The Carina-Near Moving Group}
The Carina-Near Moving Group (CARN) was identified in \cite{zuckerman2006}, and has an age estimate of 200$\pm$50 Myr \citep{zuckerman2006}.  Our search returned six known L- and T-type members or candidate members, summarized in Table \ref{tab:carn}.  No new CARN candidates were found in this search.

\begin{longrotatetable}
\begin{deluxetable*}{lccccccccccccccc}
\label{tab:carn}
\tablecaption{Carina-Near Members and Candidate Members }
\tablehead{
\colhead{Name} & \colhead{Sp.} & \colhead{Ref.} & \colhead{$\mu_{\alpha}$} & \colhead{$\mu_{\delta}$} & \colhead{Ref.} & \colhead{$\varpi$} & \colhead{Ref.} & \colhead{BANYAN} & \colhead{LACEwING} &\colhead{Ref.\tablenotemark{a}}  \\
\colhead{          } & \colhead{Type} & \colhead{      } & \colhead{(mas yr$^{-1}$)} & \colhead{(mas yr$^{-1}$)} & \colhead{ } & \colhead{(mas)} & \colhead{ } & \colhead{(\%)} & \colhead{(\%)} &\colhead{ }  }
\startdata
\cutinhead{Recovered Members}
PSO J004.6359$+$56.8370 & T4.5 & 1 & 375.9$\pm$3.2 & 10.4$\pm$2.7 & 1 & 46.5$\pm$3.9 & 1 & 89.7 & 0.0 & 2 \\
WISE J003110.04$+$574936.3 & L9 & 3 & 521.8$\pm$1.5 & -18.3$\pm$1.6 & 4 & 71.0$\pm$3.2 & 4 & 97.6 & 0.0 & 5 \\
WISE J031624.35$+$430709.1 & T8 & 6 & 375.5$\pm$0.9 & -227.4$\pm$0.9 & 4 & 74.7$\pm$2.1 & 4 & 94.4 & 0.0 & 2 \\
2MASSI J1553022$+$153236 & T6.5+T7.5 & 7 & -385.8$\pm$0.7 & 166.2$\pm$0.9 & 7 & 75.1$\pm$0.9 & 7 & 98.1 & 0.0 & 2,8 \\
SDSSp J162414.37$+$002915.6 & T6 & 9 & -372.89$\pm$1.60 & -9.11$\pm$1.95 & 10 & 90.9$\pm$1.2 & 10 & 27.0 & 0.0 & 2,8 \\
WISE J223617.59$+$510551.9 & T5 & 3 & 719.1$\pm$1.6 & 350.0$\pm$1.8 & 11 & 102.8$\pm$1.9 & 1 & 96.7 & 0.0 & 2,4,8 \\
\enddata
\tablenotetext{a}{References in the final column are the relevant works that have previously evaluated, suggested, or determined moving group membership. }
\tablerefs{(1) \cite{best2020}; (2) \cite{zhang2021}; (3) \cite{best2013}; (4) \cite{kirkpatrick2021}; (5) \cite{vos2022}; (6) \cite{mace2013}; (7) \cite{dupuy2012}; (8) \cite{hsu2021}; (9) \cite{burgasser2002}; (10) \cite{tinney2003}; (11) This work}
\end{deluxetable*}
\end{longrotatetable}

\subsubsection{The Coma Ber Cluster}

The Coma Ber Cluster (CBER) is the second closest open cluster to the Sun after the Hyades, and has recent age estimates of 700--800 Myr \citep{tang2018, martin2020, sapozhnikov2021, singh2023}.  Two CBER candidates were found in this search.  Using the Gaia DR3 proper motion and parallax of SDSS J125108.28$+$155911.1 (L0; \citealt{kiman2019}) we find a high probability of CBER membership from LACEwING (88.0\%), though BANYAN $\Sigma$ returns a 0\% probability of CBER membership.  If SDSS J125108.28$+$155911.1 is a CBER member, it would exist outside of the tidal radius.  By comparing the Cartesian XYZ coordinates of SDSS J125108.28$+$155911.1 found from Gaia DR3 astrometry (7.5 pc, -11.8 pc, 70.9 pc) to the coordinates of core and tidal tail members from \cite{tang2019}, we find that SDSS J125108.28$+$155911.1 would have the lowest Z value of all candidate members by $\sim$6.4 pc.  We suggest that CBER membership is therefore unlikely for SDSS J125108.28$+$155911.1. 

The other CBER candidate is 2MASS J13264464$+$3627407 (L2; \citealt{kellogg2017}).  Using the UHS proper motion and $J$-band photometric distance of this source (no parallax is available), LACEwING returns a 93.0\% probability of CBER membership, while BANYAN $\Sigma$ does not find a significant probability of CBER membership.  We retain 2MASS J13264464$+$3627407 as a tentative candidate requiring additional validation.  We do not flag this object as young in Table \ref{tab:props} because brown dwarfs with ages of $\sim$700 Myr are not known to be readily distinguishable from the field L and T population.

\subsubsection{The Hyades Cluster}
\label{sec:hya}

The Hyades is the closest known open cluster to the Sun, and has an age of $\sim$650 Myr (e.g., \citealt{lodieu2019}).  Twenty-seven previously suggested L- and T-type Hyades members were recovered in our search, in addition to eighteen new candidate members, summarized in Table \ref{tab:hyades}.  Of the recovered members, we highlight 2MASS J01311838$+$3801554, 2MASS J01472702$+$4731142, and 2MASSW J0208183$+$254253. 2MASS J01311838$+$3801554 and 2MASS J01472702$+$4731142 were returned as strong Hyades candidates from LACEwING, with weak probabilities from BANYAN $\Sigma$, while 2MASSW J0208183$+$254253 had a strong LACEwING probability and moderately high probability from BANYAN $\Sigma$. All three of these objects are included in the list of possible Hyades cluster members from \cite{gaia2021}.  These objects exist well outside of the core radius of the Hyades cluster, with distances of 30-38 pc from the cluster center \citep{gaia2021}, which is why these objects were missed as Hyades members in previous searches and the recent compilation of L- and T-type Hyades members in \cite{schneider2022}.

Fourteen objects (WISEA J015812.03$+$323157.9, SDSS J020608.97$+$223559.2, 2MASSW J0208236$+$273740, WISEA J022721.93$+$235654.3, 2MASSW J0242435$+$160739, UGCS J030013.86$+$490142.5, WISEPA J030533.54$+$395434.4, 2MASSW J0310599$+$164816, PSO J049.1124$+$17.0885, 2MASS J03302948$+$3910242, SIMP J03570493$+$1529270, WISEA J041743.13$+$241506.3, PSO J070.3773$+$04.7333, and 2MASS J06143818$+$3950357) were returned as high-probability Hyades members from LACEwING, but have a 0\% probability of belonging to the Hyades from BANYAN $\Sigma$.  The directions of their proper motion components do not align with known Hyades members, which is confirmed via a convergent point analysis. We find that all of these objects are unlikely to be Hyades members.

SDSS J011912.22$+$240331.6 (T2.5; \citealt{burgasser2010}) is a new candidate Hyades member identified in this work.  It is suggested as a binary candidate based on spectral decomposition  \citep{burgasser2010, bardalez2014, ashraf2022}, however, it remains unresolved \citep{radigan2013, bardalez2014}.  Using the convergent point of the Hyades from \cite{madsen2002} and following the methods in \cite{hogan2008}, we find a difference between the proper motion angle ($\theta_{\mu}$) and convergent point angle ($\theta_{\rm cp}$) of less than 0\fdg2 (angles should be similar for true members).  We also find a cluster distance ($d_{\rm c}$) using the cluster velocity of \cite{lodieu2019} of 35.9 pc, which is in excellent agreement with the measured distance of 34.6$\pm$3.5 pc from \cite{best2020}.  We therefore consider SDSS J011912.22$+$240331.6 a new T-type Hyades candidate member.

WISEPA J020625.26$+$264023.6 was discovered in \cite{kirkpatrick2011}, and was found to be a good match to the L9 standard at $J$, but much redder at $H$ and $K$.  \cite{liu2016} reclassified it as L8 (red).  We find the same type as \cite{kirkpatrick2011} in this work (L9 red).  \cite{liu2016} found no matching groups using BANYAN II \citep{gagne2014}.  However, BANYAN II did not include the Hyades and LACEwING has not previously been used for this object to our knowledge.  \cite{ashraf2022} found this object to be a strong variability candidate with a 66.9\% of belonging to the field, though it is not clear what astrometry was used for this determination.  We find a strong Hyades membership probability for this object from LACEwING, with moderate probability from BANYAN $\Sigma$. We find a difference between $\theta_{\mu}$ and $\theta_{\rm cp}$ for this object of $\sim$0\fdg6.  We also find a $d_{\rm c}$ of 20.0 pc, which agrees well with the measured distance of 19.2$\pm$0.5 pc from \cite{liu2016}.  We consider WISEPA J020625.26$+$264023.6 a new substellar candidate Hyades member.     

2MASSW J0228110$+$253738 was discovered in \cite{wilson2003}, with an uncertain near-infrared type of L0.  We find a near-infrared spectral type of L1 in this work.  This object is returned as a high-probability Hyades member from both BANYAN $\Sigma$ and LACEwING.  Further, there exists a radial velocity measurement for this object from \cite{blake2010} (23.07$\pm$0.21 \kms).  Using this radial velocity with the additional astrometry for this object listed in Table \ref{tab:hyades}, we find a 89.1\% Hyades membership probability from BANYAN $\Sigma$ and a 96\% membership probability from LACEwING.  The difference between $\theta_{\mu}$ and $\theta_{\rm cp}$ for this object is less than 0\fdg01, and the $d_{\rm c}$ of 34.5 pc aligns well with the measured distance of 33.3$\pm$0.5 pc from \cite{dahn2017}.  We consider 2MASSW J0228110$+$253738 a new candidate Hyades member. 

2MASSW J0306268$+$154514 was discovered in \cite{kirkpatrick2000} with an optical type of L6:, and we find a near-infrared type of L6.5 in this work.  This object was only just omitted by the Hyades study of \cite{schneider2022}, who had a $\mu_{\alpha}$ requirement of $\leq$197 mas yr$^{-1}$ (this object has $\mu_{\alpha}$ = 205.4 mas yr$^{-1}$).  This object just missed the probability threshold of 80\% for BANYAN $\Sigma$ (78.7\%) and has a strong Hyades membership probability from LACEwING (98\%).  This candidate does not have a measured parallax, but the predicted distances from BANYAN $\Sigma$ (36.4 pc) LACEwING (36.1 pc), and the convergent point (36.7 pc) are all generally consistent with this object's photometric distance estimates (46$\pm$8 pc from $J$, 37$\pm$6 pc from $K$). We consider 2MASSW J0306268$+$154514 a new Hyades candidate member. 

\begin{figure*}
\plotone{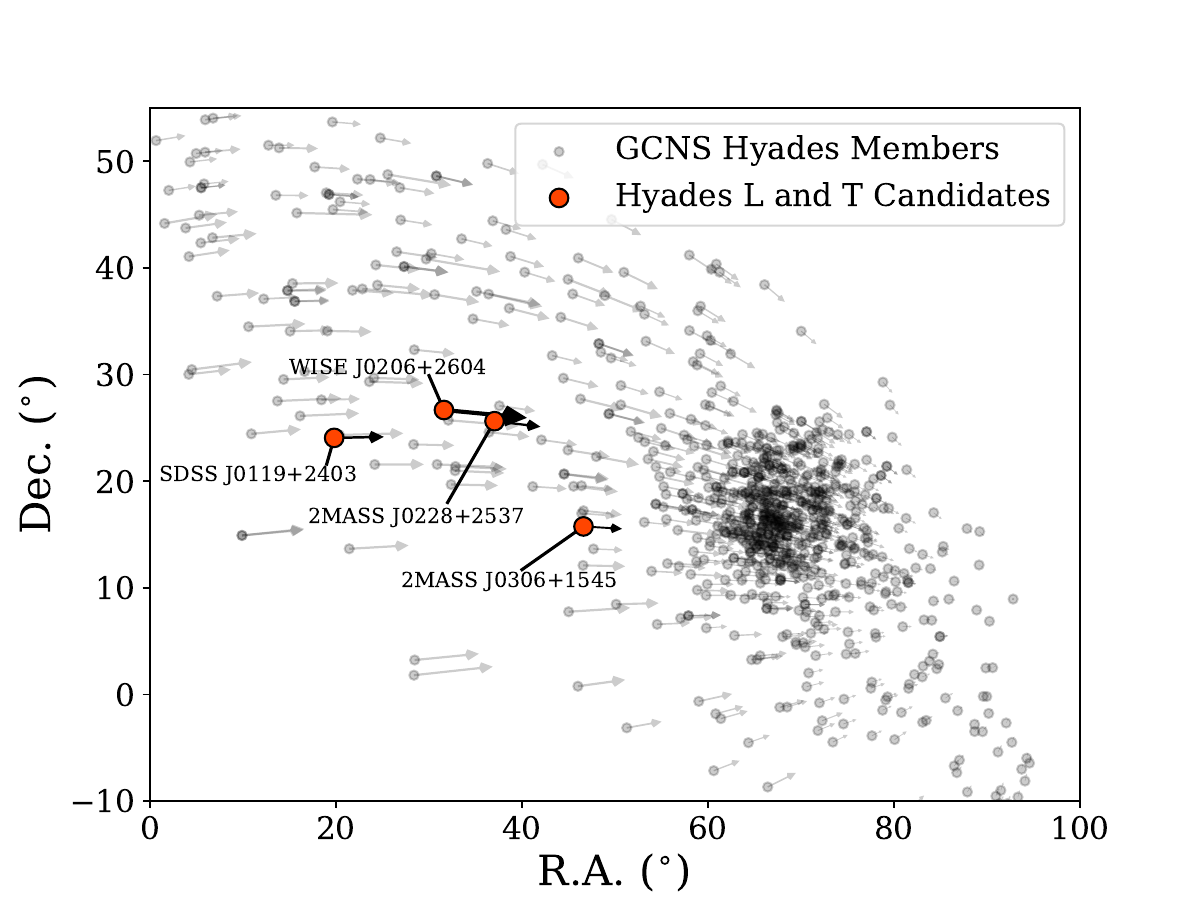}
\caption{The positions and proper motions of the four new Hyades candidate members identified in this work compared to the list of Hyades members from the Gaia Catalog of Nearby Stars (GCNS; \citealt{gaia2021}). }  
\label{fig:hyades}
\end{figure*}

The four new Hyades candidates identified in this work are shown compared to the Hyades member compilation from \cite{gaia2021} in Figure \ref{fig:hyades}.  It is clear in the figure that these four candidates exist outside of the cluster center, but have motions consistent with known Hyades members.  We suggest these four objects as new Hyades cluster candidate  members.

\begin{longrotatetable}
\begin{deluxetable*}{lccccccccccccccc}
\label{tab:hyades}
\tablecaption{Hyades Members and Candidate Members }
\tablehead{
\colhead{Name} & \colhead{Sp.} & \colhead{Ref.} & \colhead{$\mu_{\alpha}$} & \colhead{$\mu_{\delta}$} & \colhead{Ref.} & \colhead{$\varpi$\tablenotemark{a}} & \colhead{Ref.} & \colhead{BANYAN} & \colhead{LACEwING} &\colhead{Ref.\tablenotemark{b}}  \\
\colhead{          } & \colhead{Type} & \colhead{      } & \colhead{(mas yr$^{-1}$)} & \colhead{(mas yr$^{-1}$)} & \colhead{ } & \colhead{(mas)} & \colhead{ } & \colhead{(\%)} & \colhead{(\%)} &\colhead{ }  }
\startdata
\cutinhead{Recovered Members}
2MASS J01311838$+$3801554 & L1.5 & 1 & 380.05$\pm$0.45 & -33.95$\pm$0.43 & 2 & 40.92$\pm$0.50 & 2 & 2.3 & 87.0 & 3 \\ 
2MASS J01472702$+$4731142 & L1 & 4 & 184.75$\pm$1.26 & -30.13$\pm$1.16 & 2 & 20.85$\pm$1.36 & 2 & 0.0 & 82.0 & 3 \\ 
2MASSW J0208183$+$254253 & L1.5 & 4 & 374.51$\pm$0.34 & -30.51$\pm$0.34 & 2 & 43.10$\pm$0.32 & 2 & 68.7 & 95.0 & 3 \\
CWISE J031042.59$+$204629.3 & L5 & 5 & 161.2$\pm$2.0 & -27.2$\pm$2.1 & 6 & [32$\pm$6] & 7 & 47.8 & 91.0 & 6 \\
PSO J049.1159$+$26.8409 & T2.5 & 8 & 201.1$\pm$2.4 & -52.8$\pm$1.9 & 9 & 33.5$\pm$3.1 & 9 & 85.6 & 96.0 & 6,10 \\
PSO J052.2746$+$13.3754 & T3.5 & 9 & 273.2$\pm$2.0 & -20.7$\pm$2 & 9 & 44.3$\pm$3.0 & 9 & 92.5 & 96.0 & 6,10 \\
CWISE J033817.87$+$171744.1 & L7 & 6 & 152.0$\pm$3.5 & -25.3$\pm$3.3 & 6 & [22$\pm$4] & 7 & 85.4 & 100.0 & 6 \\
2MASS J03530419$+$0418193 & L6 (red) & 5 & 171.2$\pm$3.1 & 35.8$\pm$2.9 & 6 & [31$\pm$5] & 7 & 5.7 & 100.0 & 6 \\
SDSS J035308.54$+$103056.0 & L1 & 11 & 128.75$\pm$1.36 & -19.61$\pm$0.78 & 2 & 21.56$\pm$0.95 & 2  & 0.0 & 100.0 & 12 \\
Hya03 & L1 pec (red) & 13 & 110.25$\pm$1.43 & -13.06$\pm$1.34 & 2 & 17.90$\pm$1.20 & 2 & 72.6 & 100.0 & 6,12,14,15,16,17 \\  
WISEA J041232.77$+$104408.3 & L5: (red) & 18 & 129.5$\pm$3.8 & -5.5$\pm$3.5 & 6 & [16$\pm$3] & 7 & 72.9 & 97.0 & 6,18 \\
Hya10 & L1 & 16 & 123.6$\pm$2.7 & -17.8$\pm$2.3 & 6 & 28.5$\pm$3.9 & 17 & 99.2 & 100.0 & 6,14,17,17 \\
2MASS J04183483$+$2131275 & L5 & 19 & 141.5$\pm$2.7 & -45.7$\pm$2.3 & 17 & 25.8$\pm$2.9 & 17 & 98.3 & 100.0 & 6,12,17,19 \\
CWISE J041953.55$+$203628.0 & T4 & 6 & 109.4$\pm$9.0 & -35.8$\pm$8.9 & 6 & [27$\pm$5] & 7 & 97.9 & 100.0 & 6 \\
2MASS J04241856$+$0637448 & L4 & 12 & 138.8$\pm$2.8 & 7.5$\pm$2.9 & 6 & [16$\pm$3] & 7 & 20.9 & 87.0 & 6,12 \\
CWISE J042731.38$+$074344.9 & L7 & 6 & 114.3$\pm$3.5 & 5.5$\pm$3.1 & 6 & [23$\pm$4] & 7 & 96.2 & 100.0 & 6 \\
CFHT-Hy-21 & T1 & 20 & 82.1$\pm$9.8 & -15.5$\pm$8.6 & 17 & 33.5$\pm$12.7 & 17 & 74.7 & 83.0 & 16,17,20 \\
CWISE J043018.70$+$105857.1 & T4 & 6 & 106.3$\pm$6.9 & -10.7$\pm$6.9 & 6 & [25$\pm$4] & 7 & 98.7 & 100.0 & 6 \\
CFHT-Hy-20 & T2.5 & 21 & 142.6$\pm$1.6 & -16.5$\pm$1.7 & 21 & 30.8$\pm$1.5 & 21 & 98.7 & 100.0 & 6,10,17,20,21 \\
Hya12 & L6 (red) & 7 & 100.2$\pm$1.9 & -15.1$\pm$2.0 & 17 & 24.1$\pm$2.1 & 17  & 99.5 & 100.0 & 6,14,16,17 \\
WISEA J043642.75$+$190134.8 & L6 & 18 & 113.5$\pm$2.0 & -42.1$\pm$2.0 & 6 & [25$\pm$4] & 7 & 99.1 & 100.0 & 6,18 \\
PSO J069.7303$+$04.3834 & T2 & 9 & 118.7$\pm$3.5 & 11.7$\pm$3.4 & 6 & 36.6$\pm$5.7 & 9 & 86.1 & 94.0 & 6,10 \\
CWISE J043941.41$+$202514.8 & T3 & 6 & 80.8$\pm$8.0 & -30.3$\pm$7.9 & 6 & [25$\pm$4] & 7 & 88.5 & 99.0 & 6 \\
WISEA J044105.56$+$213001.5 & L5 (red) & 18 & 97.7$\pm$4.3 & -43.6$\pm$4.0 & 6 & [16$\pm$3] & 7 & 84.6 & 99.0 & 6,18 \\
Hya09 & L2 & 16 & 76.3$\pm$2.9 & -17.7$\pm$1.5 & 17 & 20.6$\pm$2.5 & 17  & 96.9 & 100.0 & 6,14,16,17 \\
Hya08 & L0.5 & 16 & 88.63$\pm$1.12 & -17.45$\pm$0.77 & 2 & 22.73$\pm$0.88 & 2  & 99.0 & 100.0 & 6,14,16,17 \\
CWISE J053204.60$+$111955.1 & L7 & 6 & 72.5$\pm$1.9 & -30.2$\pm$1.9 & 6 & [41$\pm$7] & 7 & 59.7 & 94.0 & 6 \\
\cutinhead{New Candidate Members}
SDSS J011912.22$+$240331.6 & T2.5 & 1 & 264.6$\pm$1.5 & 5.5$\pm$1.6 & 9 & 28.9$\pm$2.9 & 9 & 1.4 & 91.0 & \dots \\
WISEPA J020625.26$+$264023.6 & L9 (red) & 22 & 442.7$\pm$2.1 & -41.9$\pm$2.3 & 20 & 52.1$\pm$1.4 & 20  & 33.1 & 89.0 & \dots \\
2MASSW J0228110$+$253738 & L1 & 7 & 244.62$\pm$0.21 & -29.60$\pm$0.42 & 23 & 30.03$\pm$0.46 & 23  & 68.8 & 98.0 & \dots \\
2MASSW J0306268$+$154514 & L6.5 & 7 & 205.4$\pm$3.5 & -12.8$\pm$3.7 & 7 & [22$\pm$4] & 7 & 52.5 & 100.0 & \dots \\
\cutinhead{Rejected Candidate Members}
WISEA J015812.03$+$323157.9 & L4.5 & 24 & 353.98$\pm$2.31 & -315.51$\pm$2.05 & 7 & [31$\pm$5] & 7 & 0.0 & 87.0 & \dots \\
SDSS J020608.97$+$223559.2 & L5.5 & 25 & 396.91$\pm$2.63 & -64.69$\pm$2.36 & 7 & [28$\pm$5] & 7 & 0.0 & 86.0 & \dots \\
2MASSW J0208236$+$273740 & L6 & 4 & 206.3$\pm$2.4 & -116.1$\pm$2.8 & 7 & 21.3$\pm$2.7 & 9 & 0.0 & 93.0 & \dots \\
WISEA J022721.93$+$235654.3 & L9 & 26 & 315.13$\pm$3.34 & -127.98$\pm$3.22 & 7 & [35$\pm$6] & 7 & 0.0 & 98.0 & \dots \\
2MASSW J0242435$+$160739 & L2 (sl. blue) & 7 & 155.26$\pm$1.76 & -207.41$\pm$1.65 & 2 & 21.31$\pm$1.85 & 2  & 0.0 & 95.0 & \dots \\
UGCS J030013.86$+$490142.5 & T6.5 & 27 & 94.9$\pm$48.8 & -130.3$\pm$52.0 & 28 & [27$\pm$5] & 7 & 0.0 & 81.0 & \dots \\
WISEPA J030533.54$+$395434.4 & T6 & 22 & 273$\pm$1.5 & 4.7$\pm$1.5 & 9 & 27.8$\pm$2.0 & 9 & 0.0 & 97.0 & \dots \\
2MASSW J0310599$+$164816 & L9 (sl. red) & 7 & 242.5$\pm$2.5 & 4.1$\pm$1.9 & 2 & 36.9$\pm$3.4 & 29 & 0.0 & 98.0 & \dots \\
PSO J049.1124$+$17.0885 & L9.5 & 8 & 267.8$\pm$3.8 & -104.9$\pm$3.4 & 7 & [31$\pm$5] & 7  & 0.0 & 89.0 & \dots \\
2MASS J03302948$+$3910242 & L7 pec (red) & 5 & 67.23$\pm$3.50 & -116.79$\pm$3.64 & 7 & [21$\pm$4] & 7 & 0.0 & 87.0 & \dots \\
SIMP J03570493$+$1529270 & L1:: pec (blue) & 13 & 96.94$\pm$0.61 & -286.36$\pm$0.43 & 2 & 24.98$\pm$0.59 & 2  & 0.0 & 98.0 & \dots \\
WISEA J041743.13$+$241506.3 & T6 & 26 & 394.76$\pm$4.61 & -508.76$\pm$3.56 & 7 & [75$\pm$13] & 7 & 0.0 & 83.0 & \dots \\
PSO J070.3773$+$04.7333 & T4.5 & 8 & 212.5$\pm$3.1 & -105.7$\pm$2.9 & 9 & 23.5$\pm$4.9 & 9 & 0.0 & 95.0 & \dots \\ 
2MASS J06143818$+$3950357 & L9 & 30 & -31.9$\pm$1.7 & -264.9$\pm$1.8 & 9 & 44.0$\pm$2.6 & 9 & 0.0 & 88.0 & \dots \\
\enddata
\tablenotetext{a}{Parallaxes in square brackets are photometric. }
\tablenotetext{b}{References in the final column are the relevant works that have previously evaluated, suggested, or determined moving group membership. }
\tablerefs{(1) \cite{burgasser2010}; (2) \cite{gaia2022}; (3) \cite{gaia2021}; (4) \cite{schneider2014}; (5) \cite{kellogg2017}; (6) \cite{schneider2022}; (7) This work; (8) \cite{best2015}; (9) \cite{best2020}; (10) \cite{zhang2021}; (11) \cite{bardalez2014}; (12) \cite{perez2018}; (13) \cite{robert2016}; (14) \cite{hogan2008}; (15) \cite{casewell2014}; (16) \cite{lodieu2014}; (17) \cite{lodieu2019}; (18) \cite{schneider2017};  (19) \cite{perez2017}; (20) \cite{bouvier2008};(20) \cite{liu2016};  (22) \cite{kirkpatrick2011}; (23) \cite{dahn2017}; (24) \cite{kirkpatrick2016}; (25) \cite{chiu2006};  (26) \cite{greco2019}; (27) \cite{lodieu2009}; (28) \cite{marocco2020}; (29) \cite{smart2013}; (30) \cite{muzic2012};  }
\end{deluxetable*}
\end{longrotatetable}

\subsubsection{The $\mu$ Tau Association}
The $\mu$ Tau Association (MUTA) was first identified in \cite{gagne2020}, with an estimated age of 62$\pm$7 Myr.  Our search returned a single candidate MUTA member, SIMP J03314657$+$1944246.  This object was discovered in \cite{robert2016}, who gave a spectral type of L0:: from a very low-S/N near infrared spectrum.  Using UHS proper motions, we find a 97\% chance of belonging to MUTA from BANYAN $\Sigma$ (note that MUTA is not included in LACEwING).  The predicted distance if a $\mu$ Tau member is $\sim$146 pc, which is not a great match to the photometric distance estimates for this object of 85$\pm$14 and 91$\pm$15 pc for $J$- and $K$-band photometry.  However, considering the very uncertain nature of this object's spectral type, it could indeed be a member if it's actual spectral type is earlier than L0.  A higher-S/N spectrum of this object would help to clear up potential MUTA membership. 

\subsubsection{The Taurus Association}

Our search recovered one known Taurus member (2MASS J04373705$+$2331080).  Our UHS proper motion is an improvement over previous measurements, and we find a 98.7\% chance of belonging to Taurus according to BANYAN $\Sigma$ (Taurus is not evaluated in LACEwING).  This known Taurus member was discovered in \cite{luhman2009}, and is further discussed in \citep{best2017, esplin2017, kraus2017}.  

\subsubsection{The 32 Ori Group}
The 32 Ori Group (THOR) was first identified in \citep{mamajek2007} with an age estimate of 15-20 Myr \citep{luhman2022}.  One high probability THOR member was recovered in our search, the known member WISE J052857.68$+$090104.4 (L1 VL-G; \citealt{burgasser2016}).  The UHS proper motion of this source is the most precise measurement available, and we find a 99.1\% probability of THOR membership from BANYAN $\Sigma$ and 97.0\% probability of THOR membership from LACEwING.  

\subsubsection{The Ursa Major Association}
The Ursa Major Association (UMA) is a loose collection of nearby stars first identified in \cite{eggen1992} with an estimated age of 414$\pm$23 \citep{jones2015}.  Our search returned four previously suggested UMA members and two new candidate members, summarized in Table \ref{tab:ursa}. 

The two new UMA candidates, SDSS J135852.68$+$374711.9 and PSO J224.3820$+$47.4057, both have high probabilities of UMA membership from LACEwING using proper motions and parallaxes from \cite{best2020}, but return a 0\% probability of UMA membership from BANYAN $\Sigma$.  We retain both objects as UMA candidate members.

\begin{longrotatetable}
\begin{deluxetable*}{lccccccccccccccc}
\label{tab:ursa}
\tablecaption{Ursa Major Members and Candidate Members }
\tablehead{
\colhead{Name} & \colhead{Sp.} & \colhead{Ref.} & \colhead{$\mu_{\alpha}$} & \colhead{$\mu_{\delta}$} & \colhead{Ref.} & \colhead{$\varpi$} & \colhead{Ref.} & \colhead{BANYAN} & \colhead{LACEwING} &\colhead{Ref.\tablenotemark{a}}  \\
\colhead{          } & \colhead{Type} & \colhead{      } & \colhead{(mas yr$^{-1}$)} & \colhead{(mas yr$^{-1}$)} & \colhead{ } & \colhead{(mas)} & \colhead{ } & \colhead{(\%)} & \colhead{(\%)} &\colhead{ }  }
\startdata
\cutinhead{Recovered Members}
WISEA J120104.57$+$573004.2 & L9 & 1 & 99.0$\pm$3.5 & 13.2$\pm$3.6 & 2 & [28$\pm$5] & 2 & 91.1 & 0.0 & 3 \\
2MASSW J1239272$+$551537 & L5 & 4 & 125.2$\pm$1.1 & -0.4$\pm$1.1 & 5 & 42.4$\pm$2.1 & 5 & 94.1 & 100.0 & 6 \\
2MASSW J1246467$+$402715 & L5 & 2 & 129.57$\pm$0.47 & -103.97$\pm$0.46 & 7 & 44.74$\pm$0.63 & 7 & 0.0 & 98.0 & 6,8 \\
SDSS J125011.65$+$392553.9 & T4 & 9 & -42.4$\pm$3.4 & -830.5$\pm$2.6 & 10 & 42.8$\pm$3.2 & 10 & 0.0 & 100.0 & 11 \\
\cutinhead{New Candidate Members}
SDSS J135852.68$+$374711.9 & T5 & 12 & -27.0$\pm$2.8 & -455.7$\pm$2.5 & 10 & 49.6$\pm$3.1 & 10 & 0.0 & 100.0 & \dots \\
PSO J224.3820$+$47.4057 & T7 & 13 & 140.0$\pm$3.0 & -84.7$\pm$2.3 & 10 & 49.5$\pm$2.9 & 10 & 0.0 & 99.0 & \dots \\
\enddata
\tablenotetext{a}{References in the final column are the relevant works that have previously evaluated, suggested, or determined moving group membership. }
\tablerefs{(1) \cite{schneider2017}; (2) This work; (3) \cite{ashraf2022}; (4) \cite{schneider2014}; (5) \cite{dupuy2012}; (6) \cite{jameson2008}; (7) \cite{gaia2022}; (8) \cite{ujjwal2020}; (9) \cite{chiu2006}; (10) \cite{best2020}; (11) \cite{zhang2021}; (12) \cite{burgasser2010}; (13) \cite{best2015}}
\end{deluxetable*}
\end{longrotatetable}

\subsubsection{Objects with Ambiguous Membership}

The L dwarf 2MASSI J0103320$+$193536 was once used as the near-infrared L7 standard \citep{kirkpatrick2010}.  However, \cite{faherty2012} and \cite{allers2013} revised the near-infrared spectral type of this object as L6$\beta$ and L6 INT-G, respectively.  This object was suggested as a strong Argus candidate member in \cite{gagne2014}, however, the astrometry used for this source from \cite{faherty2012} had relatively low precision.  \cite{gagne2015} again revised the near-infrared type of this object to L6 pec, stating that it showed no clear signs of low-gravity.  \cite{faherty2016} re-evaluated membership and determined its status to be ambiguous.  We find similar results in this work using the UHS determined proper motion and a parallax from \cite{kirkpatrick2021}.  BANYAN $\Sigma$ returns a relatively small but not insignificant chance of belonging to Carina-Near (40.7\%), while LACEwING suggests Hyades membership (84.0\%) and a lower probability of ABDMG membership (56.0\%).  The membership status of this object remains ambiguous.

\subsection{High \vtan\ Objects}
\label{sec:sds}

Space velocities can be indicative of different populations of the Galaxy.  The tangential velocity (\vtan) is often used as a rough discriminator between different Galactic components (i.e., thin disk, thick disk, and halo).  We have calculated \vtan\ values for every object in our sample, using distances determined from parallaxes or $J$-band photometry and the most precise proper motions available in an effort to identify previously unknown objects with unusual kinematics.

\cite{nissen2004} suggest a \vtan\ range for thick disk objects between 85 and 180 \kms, with anything over 180 \kms\ being a potential halo member.  These values are similar to those found in \cite{torres2019} for the Gaia white dwarf population, who found 90 \kms\ $\leq$ \vtan\ $\leq$ 200 \kms\ for the thick disc population and \vtan\ $>$ 200 \kms\ for the halo (though there is substantial overlap between the different components).  We follow the criteria outlined in \cite{dupuy2012} to identify objects with a $\geq$0.5 probability of belonging to the thick disk or halo (\vtan\ $>$ 77 + 35exp(0.028$\sigma_{\vtan}$).     

Table \ref{tab:vtan} lists the thirteen objects in this sample that were found to have \vtan\ values that satisfy the \cite{dupuy2012} criteria.  Eleven objects in this sample have measured parallaxes, while the distances to the other two are based on $J$-band photometry and spectral types.   Seven known subdwarfs are recovered in this sample, three of which are newly classified in Table \ref{tab:sdspts} in this work, as well as one ``blue'' L dwarf.   

\begin{deluxetable*}{ccccccccc}
\label{tab:vtan}
\tablecaption{High \vtan\ Objects}
\tablehead{
\colhead{Name} & \colhead{Spectral} & \colhead{Ref.} & \colhead{Dist.\tablenotemark{a}} & \colhead{Ref.} & \colhead{$\mu_{\rm tot}$} & \colhead{Ref.}  & \colhead{\vtan\tablenotemark{a}} \\
\colhead{          } & \colhead{Type} & \colhead{      } & \colhead{(pc)} & \colhead{      } & \colhead{(mas yr$^{-1}$)} & \colhead{      } & \colhead{(\kms)} }
\startdata
WISEA J004326.26$+$222124.0 & sdL1 & 1 & 66.71$\pm$ 1.30 & 2 & 441.33$\pm$0.18 & 2 & 139.6$\pm$2.7 \\
WISEA J030845.36$+$325923.1 & sdL1 & 3 & 53.77$\pm$4.13 & 2 & 493.09$\pm$1.09 & 2 & 125.7$\pm$9.7 \\
SDSS J033456.32$+$010618.7 &  L0.5 & 4 & [116$\pm$20] & 3 & 401.68$\pm$5.22 & 4 & [220$\pm$38] \\
WISEA J043535.82$+$211508.9 & sdL0 & 1 & 59.71$\pm$2.12 & 2 & 1288.77$\pm$0.39 & 2 & 364.8$\pm$12.9 \\
SDSS J100016.92$+$321829.4 & L1 & 5 & [69$\pm$12] & 3 & 487.45$\pm$3.71 & 6 & [159$\pm$28] \\
SDSS J103143.09$+$524558.7 & L0 & 7 & 76.89$\pm$4.59 & 2 & 349.53$\pm$0.47 & 2 & 127.4$\pm$7.6 \\
SDSS J125045.66$+$441853.7 & L0 & 8 & 136.20$\pm$15.03 & 2 & 284.68$\pm$0.45 & 2 & 183.8$\pm$20.3 \\
2MASSW J1411175$+$393636 & L1.5 & 9 & 29.65$\pm$0.39 & 2 & 930.68$\pm$0.24 & 2 & 130.8$\pm$1.7 \\
2MASS J14343616$+$2202463 & d/sdL1 & 3 & 31.64$\pm$0.80 & 2 & 770.25$\pm$0.35 & 2 & 115.5$\pm$2.9 \\
2MASS J16262034$+$3925190 & sdL4 & 10 & 30.93$\pm$0.14 & 2 & 1395.11$\pm$0.11 & 2 & 204.5$\pm$0.9 \\
2MASS J16403197$+$1231068 & sdL0: & 3 & 101.81$\pm$4.30 & 2 & 260.60$\pm$0.24 & 2 & 125.8$\pm$5.3 \\
2MASSI J1721039$+$334415 & L3 (blue) & 3 & 16.18$\pm$0.04 & 2 & 1947.39$\pm$0.11 & 2 & 149.4$\pm$0.4 \\
2MASS J17561080$+$2815238 & sdL1 & 10 & 34.56$\pm$0.32 & 2 & 740.78$\pm$0.19 & 2 & 121.3$\pm$1.1 \\
\enddata
\tablenotetext{a}{Distances listed in square brackets are photometric $J$-band distances using Table \ref{tab:coeffs}.  Likewise, \vtan\ values given in square brackets are derived from these distances.}
\tablerefs{(1) \cite{kirkpatrick2014}; (2) \cite{gaia2022}; (3) This work.; (4) \cite{scholz2009}; (5) \cite{schmidt2010}; (6) \cite{best2018}; (7) \cite{kiman2019}; (8) \cite{west2008}; (9) \cite{bardalez2014}; (10) \cite{greco2019} }
\end{deluxetable*}

\subsection{Co-moving Companions}
\label{sec:comps}

Co-moving L, T, and Y type companions serve as valuable benchmarks for tests of stellar and substellar models (e.g., \citealt{faherty2010, deacon2014}).  We have cross-matched our sample against Gaia DR3 \cite{gaia2022} to identify any primary components of previously unrecognized multiple systems.  This search returned seven new wide multiple systems containing at least one member with a spectral type of L0 or later.  We evaluated each newly identified system with the \texttt{COMOVER} \citep{gagne2021} tool.  Photometric distances were used in \texttt{COMOVER} when parallactic distances were not available.  The new systems identified in this work, and their co-moving probabilities according to \texttt{COMOVER} are listed in Table \ref{tab:comovers}.

For SDSS J081132.87$+$485532.9 (L0; \citealt{kiman2019}), we find a white dwarf co-moving companion in Gaia, with a separation of $\sim$2\arcsec.  The UHS proper motion of SDSS J081132.87$+$485532.9 matches well with the white dwarf, but the distance from the white dwarf parallax (196.53$\pm$26.52; \citealt{gaia2022}) does not match the $J$-band photometric distance of SDSS J081132.87$+$485532.9 (106$\pm$18 pc).  It is possible that the spectral type of SDSS J081132.87$+$485532.9 determined in \cite{kiman2019} is inaccurate, and this is instead an M-type companion to a white dwarf at $\sim$200 pc.

\begin{deluxetable*}{lccccccccccccccc}
\label{tab:comovers}
\tablecaption{New Co-moving Companions}
\tablehead{
\colhead{Name} & \colhead{Sp.} & \colhead{Ref.} & \colhead{$\mu_{\alpha}$} & \colhead{$\mu_{\delta}$} & \colhead{Ref.} & \colhead{Dist.\tablenotemark{a}} & \colhead{Ref.} & \colhead{Sep.} & \colhead{\texttt{COMOVER}}   \\
\colhead{          } & \colhead{Type} & \colhead{      } & \colhead{(mas yr$^{-1}$)} & \colhead{(mas yr$^{-1}$)} & \colhead{ } & \colhead{(pc)} & \colhead{ } & \colhead{(\arcsec)} & \colhead{(\%)} }
\startdata
SDSS J075259.47$+$413646.7 & M7 & 1 & -23.21$\pm$0.18 & 10.64$\pm$0.15 & 2 & 85.22$\pm$1.15 & 2 & \dots & \dots \\
SDSS J075259.43$+$413634.6 & L0 & 3 & -21.7$\pm$2.2 & 10.1$\pm$2.3 & 4 & [95$\pm$16] & 4 & 12.0 & 99.98 \\
\hline
Gaia DR3 931634554610889728 & WD & 5 & 9.26$\pm$0.56 & -47.68$\pm$0.51 & 2 & 196.53$\pm$26.52 & 2 & \dots & \dots \\
SDSS J081132.87$+$485532.9 & L0 & 6 & 7.8$\pm$3.2 & -45.8$\pm$2.6 & 4 & [106$\pm$18] & 4 & 2.3 & 93.68 \\
\hline
TYC 813-243-1A & K5 & 7 & -41.75$\pm$0.02 & -70.55$\pm$0.01 & 2 & 64.87$\pm$0.08 & 2 & \dots & \dots \\
TYC 813-243-1B & \dots & \dots & -39.19$\pm$0.02 & -71.83$\pm$0.02 & 2 & 64.92$\pm$0.10 & 2 & 1.9 & \dots \\
SDSS J084457.38$+$120825.4 & L0 & 6 & -39.85$\pm$0.48 & -70.61$\pm$0.34 & 2 & 65.90$\pm$1.79 & 2 & 55.5 & 100.00 \\
\hline
HD 76945 & F2 & 8 & -95.66$\pm$0.06 & -26.86$\pm$0.06 & 2 & 75.13$\pm$0.35 & 2 & \dots & \dots \\
PSO J135.0395$+$32.0845 & L1.5 & 9 & -97.25$\pm$1.57 & -25.08$\pm$1.37 & 2 & 69.52$\pm$5.92 & 2 & 61.2 & 100.00 \\
\hline
LSPM J1230$+$4048 & M3 & 11 & -155.15$\pm$0.02 & 32.16$\pm$0.02 & 2 & 62.50$\pm$0.09 & 2 & \dots & \dots \\
SDSS J123112.97$+$405027.9 & L0 & 10 & -156.27$\pm$0.35 & 31.53$\pm$0.45 & 2 & 60.93$\pm$1.82 & 2 & 187.4 & 100.00 \\
\hline
ATO J199.1400$+$57.6089 & M4 & 11 & 43.12$\pm$0.09 & 19.70$\pm$0.09 & 2 & 48.08$\pm$0.22 & 2 & \dots & \dots \\
SDSS J131633.79$+$573549.1 & L0 & 6 & 48.52$\pm$0.21 & 17.49$\pm$0.20 & 2 & 48.30$\pm$0.47 & 2 & 43.0 & 100.00 \\
\hline
Gaia DR3 2816131994957537280 & \dots & \dots & 38.29$\pm$0.11 & -13.06$\pm$0.09 & 2 & 96.53$\pm$0.97 & 2 & \dots & \dots \\
SDSS J230134.21$+$144219.6 & L0 & 6 & 42.3$\pm$3.2 & -12.51$\pm$4.0 & 4 & [90$\pm$15] & 4 & 52.5 & 99.99 \\
\enddata
\tablenotetext{a}{Distances listed in square brackets are $J$-band photometric distances derived using the relation derived in Section \ref{sec:phot}.}
\tablerefs{(1) \cite{west2011}; (2) \cite{gaia2022}; (3) \cite{hawley2002}; (4) This work; (5) \cite{gentile2019}; (6) \cite{kiman2019}; (7) \cite{pickles2010}; (8) \cite{cannon1919}; (9) \cite{best2015}; (10) \cite{schmidt2010}; (11) \cite{luo2015};    }
\end{deluxetable*}

\section{Summary}
\label{sec:summary}

We have presented photometry and astrometry for all spectroscopically confirmed L, T, and Y dwarfs in the UKIRT Hemisphere Survey.  We have determined typical colors and absolute magnitudes using UHS photometry for normal L, T, and Y dwarfs.  We have also determined proper motions for each object detected in both the UHS $J$- and $K$-band surveys (768 total), 381 of which are the most precise proper motions yet measured.  

Using the best available astrometry for each object, we have strengthened the membership status for several previously suggested moving group members and identified sixtenn new candidate members. We have further identified a number of high tangential-velocity objects and seven previously overlooked wide-separation L-type co-moving companions.  

Transforming UHS positions to the Gaia reference frame vastly improved their accuracy and precision.  Similar improvements could be applied to other previous publicly available surveys.  For example, astrometric calibrations for all previous UKIDSS surveys were performed using the 2MASS Point Source Catalog \citep{dye2006}, which is itself tied to the Tycho astrometric calibration \citep{hog2000}.  Astrometric calibrations for all VISTA surveys \citep{mcmahon2013} are also performed using the 2MASS point source catalog\footnote{https://www.eso.org/rm/api/v1/public/releaseDescriptions/144}.  Re-registering these surveys would allow for new, more accurate and precise proper motion investigations, and would produce an astrometric catalog with similar near-infrared depths between declinations of $-$90$\degr$ to $+$60$\degr$, with only the north celestial pole yet to be surveyed.
 
\clearpage
\begin{acknowledgments}

This publication makes use of data products from the UKIRT Hemisphere Survey, which is a joint project of the United States Naval Observatory, The University of Hawaii Institute for Astronomy, the Cambridge University Cambridge Astronomy Survey Unit, and the University of Edinburgh Wide-Field Astronomy Unit (WFAU).  This project was primarily funded by the United States Navy.  The WFAU gratefully acknowledges support for this work from the Science and Technology Facilities Council through ST/T002956/1 and previous grants. 

This publication makes use of data products from the {\it Wide-field Infrared Survey Explorer}, which is a joint project of the University of California, Los Angeles, and the Jet Propulsion Laboratory/California Institute of Technology, and NEOWISE which is a project of the Jet Propulsion Laboratory/California Institute of Technology. {\it WISE} and NEOWISE are funded by the National Aeronautics and Space Administration.  

This work has made use of data from the European Space Agency (ESA) mission {\it Gaia} (\url{https://www.cosmos.esa.int/gaia}), processed by the {\it Gaia} Data Processing and Analysis Consortium (DPAC, \url{https://www.cosmos.esa.int/web/gaia/dpac/consortium}). Funding for the DPAC has been provided by national institutions, in particular the institutions participating in the {\it Gaia} Multilateral Agreement.  

The Pan-STARRS1 Surveys (PS1) and the PS1 public science archive have been made possible through contributions by the Institute for Astronomy, the University of Hawaii, the Pan-STARRS Project Office, the Max-Planck Society and its participating institutes, the Max Planck Institute for Astronomy, Heidelberg and the Max Planck Institute for Extraterrestrial Physics, Garching, The Johns Hopkins University, Durham University, the University of Edinburgh, the Queen's University Belfast, the Harvard-Smithsonian Center for Astrophysics, the Las Cumbres Observatory Global Telescope Network Incorporated, the National Central University of Taiwan, the Space Telescope Science Institute, the National Aeronautics and Space Administration under Grant No. NNX08AR22G issued through the Planetary Science Division of the NASA Science Mission Directorate, the National Science Foundation Grant No. AST-1238877, the University of Maryland, Eotvos Lorand University (ELTE), the Los Alamos National Laboratory, and the Gordon and Betty Moore Foundation.

This work has benefited from The UltracoolSheet at http:// bit.ly/UltracoolSheet, maintained by Will Best, Trent Dupuy, Michael Liu, Rob Siverd, and Zhoujian Zhang, and developed from compilations by \cite{dupuy2012}, \cite{dupuy2013}, \cite{liu2016}, and \cite{best2018, best2021}.

\end{acknowledgments}

\facilities{UKIRT}

\software{BANYAN~$\Sigma$ \citep{gagne2018}; LACEwING \citep{riedel2017}; COMOVER \citep{gagne2021}}

\restartappendixnumbering

\clearpage
\appendix
\section{The UHS L, T, and Y Dwarf Catalog}

The following table includes astrometry and photometry for all known L, T, and Y dwarfs in the UHS survey.  

\startlongtable
\begin{deluxetable*}{llccccccc}
\label{tab:props}
\tablecaption{Sample Properties }
\tablehead{
\colhead{Column Label} & \colhead{Description}  & \colhead{Example} & Units  \\}
\startdata
Disc\_Name & Discovery name & SDSS J000013.54$+$255418.6 & \dots  \\
Disc\_ref & Discovery reference & 1 & \dots \\
SpT\_Opt & Optical spectral type & T5 & \dots \\
SpT\_Opt\_ref & Reference for optical spectral type & 111 & \dots \\
SpT\_IR & Near Infrared spectral type & T4.5 & \dots \\
SpT\_IR\_ref & Reference for near infrared spectral type & 1 & \dots \\
RA\_UHS & UHS DR2 right ascension & 0.0563546 & degrees \\
Dec\_UHS & UHS DR2 declination & 25.9055694 & degrees \\
RA\_UHS\_J & UHS $J$-band right ascension & 0.0564025 & degrees \\
eRA\_UHS\_J & Uncertainty of UHS $J$-band right ascension & 15.10 & mas \\
Dec\_UHS\_J & UHS $J$-band declination & 25.9055032 & degrees \\
eDec\_UHS\_J & Uncertainty of UHS $J$-band declination & 12.13 & mas \\
UHS\_J\_epoch & UHS $J$-band epoch & 2013.484 & years \\
RA\_UHS\_K & UHS $K$-band right ascension & 0.5635380 & degrees \\
eRA\_UHS\_K & Uncertainty of UHS $K$-band right ascension & 10.29 & mas \\
Dec\_UHS\_K & UHS $K$-band declination & 25.9056502 & degrees \\
eDec\_UHS\_K & Uncertainty of UHS $K$-band declination & 10.76 & mas \\
UHS\_K\_epoch & UHS $K$-band epoch & 2017.749 & years \\
Jmag & UHS J magnitude & 14.8459 & mag \\
eJmag & Uncertainty of UHS J magnitude & 0.0049 & mag \\
Kmag & UHS K magnitude & 14.9702 & mag \\
eKmag & Uncertainty of UHS K magnitude & 0.0164 & mag \\
UHS\_pmra & UHS $\mu$$_{\alpha}$ & $-$26.68 & mas yr$^{-1}$ \\
UHS\_epmra & UHS uncertainty of $\mu$$_{\alpha}$ & 2.25 & mas yr$^{-1}$ \\
UHS\_pmdec & UHS $\mu$$_{\delta}$ & 127.84 & mas yr$^{-1}$ \\ 
UHS\_epmdec & UHS uncertainty of $\mu$$_{\delta}$ & 2.35 & mas yr$^{-1}$ \\
RA\_Gaia & Gaia DR3 right ascension & 0.05635442742 & degrees \\
Dec\_Gaia & Gaia DR3 declination & 25.90556949064 & degrees \\
Gaia\_plx & Parallax from Gaia DR3 & \dots & mas \\
Gaia\_eplx & Uncertainty of Gaia DR3 parallax & \dots & mas \\
Gaia\_pmra & Gaia DR3 $\mu$$_{\alpha}$ & \dots & mas yr$^{-1}$ \\
Gaia\_epmra & Gaia DR3 uncertainty of $\mu$$_{\alpha}$ & \dots & mas yr$^{-1}$ \\
Gaia\_pmdec & Gaia DR3 $\mu$$_{\delta}$ & \dots & mas yr$^{-1}$ \\ 
Gaia\_epmdec & Gaia DR3 uncertainty of $\mu$$_{\delta}$ & \dots & mas yr$^{-1}$ \\
Gmag & Gaia DR3 G magnitude & 21.380291 & mag \\
RPmag & Gaia DR3 G$_{\rm RP}$ magnitude & 19.64465 & mag \\
BPmag & Gaia DR3 G$_{\rm BP}$ magnitude & \dots & mag \\
Name\_PS1 & Pan-Starrs DR2 designation & PSO J000.0564$+$25.9054 & \dots \\
RA\_PS1 & Pan-Starrs DR2 right ascension & 0.05638003 & degrees \\
Dec\_PS1 & Pan-Starrs DR2 declination & 25.90545405 & degrees \\
gmag & Pan-Starrs $g$ magnitude & \dots & mag \\
egmag & Pan-Starrs $g$ uncertainty & \dots & mag \\
rmag & Pan-Starrs $r$ magnitude & \dots & mag \\
ermag & Pan-Starrs $r$ uncertainty & \dots & mag \\
imag & Pan-Starrs $i$ magnitude & \dots & mag \\
eimag & Pan-Starrs $i$ uncertainty & \dots & mag \\
zmag & Pan-Starrs $z$ magnitude & 19.1981 & mag \\
ezmag & Pan-Starrs $z$ uncertainty & 0.0152 & mag \\
ymag & Pan-Starrs $y$ magnitude & 17.4245 & mag \\
eymag & Pan-Starrs $y$ uncertainty & 0.0083 & mag \\
PS1\_pmra & \cite{best2018} $\mu$$_{\alpha}$ & $-$18.4 & mas yr$^{-1}$ \\
PS1\_epmra & \cite{best2018} uncertainty of $\mu$$_{\alpha}$ & 5.5 & mas yr$^{-1}$ \\
PS1\_pmdec & \cite{best2018} $\mu$$_{\delta}$ & 123.1 & mas yr$^{-1}$ \\ 
PS1\_epmdec & \cite{best2018} uncertainty of $\mu$$_{\delta}$ & 3.3 & mas yr$^{-1}$ \\
Name\_CWISE & CatWISE 2020 designation & J000013.51$+$255419.7 & \dots \\
RA\_CWISE & CatWISE 2020 right ascension & 0.0563043 & degrees \\
Dec\_CWISE & CatWISE 2020 declination & 25.9054854 & degrees \\
W1 & CatWISE 2020 W1 magnitude & 12.922 & mag \\
eW1 & CatWISE 2020 W1 magnitude uncertainty & 0.017 & mag \\
W2 & CatWISE 2020 W2 magnitude & 12.155 & mag \\
eW2 & CatWISE 2020 W2 magnitude uncertainty & 0.010 & mag \\
CWISE\_pmra & CatWISE 2020 $\mu$$_{\alpha}$ & $-$20.31 & mas yr$^{-1}$ \\
CWISE\_epmra & CatWISE 2020 uncertainty of $\mu$$_{\alpha}$ & 6.4 & mas yr$^{-1}$ \\
CWISE\_pmdec & CatWISE 2020 $\mu$$_{\delta}$ & 50.97 & mas yr$^{-1}$ \\ 
CWISE\_epmdec & CatWISE 2020 uncertainty of $\mu$$_{\delta}$ & 6.3 & mas yr$^{-1}$ \\
O\_pmra & Optimal $\mu$$_{\alpha}$ & $-$19.09 & mas yr$^{-1}$ \\
O\_epmra & Uncertainty of optimal $\mu$$_{\alpha}$ & 1.44 & mas yr$^{-1}$ \\
O\_pmdec & Optimal $\mu$$_{\delta}$ & 126.67 & mas yr$^{-1}$ \\
O\_epmdec & Uncertainty of optimal $\mu$$_{\delta}$ & 1.30 & mas yr$^{-1}$ \\
pm\_ref & Reference for optimal proper motion & 132 & \dots \\
O\_plx & Optimal parallax & 70.8 & mas \\
O\_eplx & Uncertainty of optimal parallax & 1.9 & mas \\
plx\_ref & Reference for optimal parallax & 132 & \dots \\
flag & flag\tablenotemark{a} & 10,14 & \dots \\
\enddata
\tablecomments{This table is available in its entirety in a machine-readable form in the online journal.}
\tablenotetext{a}{Flags: 1 = young or suggested to be young; 2 = subdwarf; 3 = visual binary; 4 = Hyades member or candidate member; 5 = red near-infrared spectrum; 6 = spectral binary; 7 = wide companion; 8 = peculiar spectrum; 9 = uncertain spectral type; 10 = poor or contaminated CatWISE 2020 photometry; 11 = poor or contaminated Pan-STARRS photometry; 12 = blue near-infrared spectrum; 13 = Gaia RUWE $\geq$ 1.4; 14 = photometry from CatWISE 2020 reject catalog; 15 = Parallax given is from co-moving companion; 16 = UHS photometry near saturation limit}
\tablerefs{(1) \cite{knapp2004}; (2) \cite{chiu2006}; (3) \cite{mace2013}; (4) \cite{zhang2010}; (5) \cite{schneider2016}; (6) \cite{kirkpatrick2010}; (7) \cite{kellogg2017}; (8) \cite{kirkpatrick2000}; (9) \cite{best2015}; (10) \cite{best2020}; (11) \cite{aller2016}; (12) \cite{zhang2019}; (13) \cite{reid2006}; (14) \cite{burgasser2010}; (15) \cite{deacon2014}; (16) \cite{kirkpatrick1999}; (17) \cite{thompson2013}; (18) \cite{reid2000}; (19) \cite{mugrauer2006}; (20) \cite{burgasser2004}; (21) \cite{luhman2014}; (22) \cite{wilson2003}; (23) \cite{meisner2020}; (24) \cite{naud2014}; (25) \cite{kirkpatrick2011}; (26) \cite{cruz2007}; (27) \cite{kirkpatrick2012}; (28) \cite{robert2016}; (29) \cite{schneider2017}; (30) \cite{dayjones2013}; (31) \cite{hawley2002}; (32) \cite{schmidt2010}; (33) \cite{deacon2017}; (34) \cite{cruz2003}; (35) \cite{albert2011}; (36) \cite{scholz2011}; (37) \cite{lodieu2009}; (38) \cite{luhman2014b}; (39) \cite{schneider2022}; (40) \cite{marocco2015}; (41) \cite{west2008}; (42) \cite{gizis2003}; (43) \cite{west2011}; (44) \cite{vos2022}; (45) \cite{zhang2020}; (46) \cite{reid2008}; (47) \cite{cushing2011}; (48) \cite{hogan2008}; (49) \cite{perez2017}; (50) \cite{luhman2006}; (51) \cite{perez2018}; (52) \cite{bouvier2008}; (53) \cite{luhman2009}; (54) \cite{schneider2023}; (55) \cite{bihain2013}; (56) \cite{looper2007}; (57) \cite{lepine2002}; (58) \cite{kirkpatrick2021}; (59) \cite{muzic2012}; (60) \cite{best2013}; (61) \cite{kirkpatrick2014}; (62) \cite{thorstensen2003}; (63) \cite{burgasser2002}; (64) \cite{kiman2019}; (65) \cite{zhang2009}; (66) \cite{skrzypek2016}; (67) \cite{reyle2018}; (68) \cite{smith2014}; (69) \cite{schmidt2015}; (70) \cite{scholz2010}; (71) \cite{aberasturi2011}; (72) \cite{geballe2002}; (73) \cite{deacon2009}; (74) \cite{bouy2003}; (75) \cite{reyle2010}; (76) \cite{wilson2001}; (77) \cite{burningham2013}; (78) \cite{cushing2014}; (79) \cite{tinney1993}; (80) \cite{sheppard2009}; (81) \cite{rebolo1998}; (82) \cite{baron2015}; (83) \cite{schmidt2007}; (84) \cite{burgasser1999}; (85) \cite{kuchner2017}; (86) \cite{wright2013}; (87) \cite{kellogg2015}; (88) \cite{gizis2000}; (89) \cite{deacon2011}; (90) \cite{stern2007}; (91) \cite{burgasser2003}; (92) \cite{faherty2009}; (93) \cite{metchev2008}; (94) \cite{kirkpatrick2016}; (95) \cite{burgasser2011}; (96) \cite{artigau2011}; (97) \cite{strauss1999}; (98) \cite{bardalez2014}; (99) \cite{schneider2002}; (100) \cite{radigan2008}; (101) \cite{gizis2011}; (102) \cite{marocco2020}; (103) \cite{luhman2012}; (104) \cite{looper2008}; (105) \cite{jameson2008}; (106) \cite{gizis2011b}; (107) \cite{burgasser2006}; (108) \cite{luhman2007}; (109) \cite{martin2018}; (110) \cite{dahn2002}; (111) \cite{pineda2016}; (112) \cite{zhang2017}; (113) \cite{cruz2009}; (114) \cite{scholz2009}; (115) \cite{lodieu2014}; (116) \cite{burgasser2003c}; (117) \cite{faherty2013}; (118) \cite{burgasser2007}; (119) \cite{cruz2018}; (120) \cite{gizis2015}; (121) \cite{gizis2013}; (122) \cite{kirkpatrick2008}; (123) This work; (124) \cite{schneider2014}; (125) \cite{allers2013}; (126) \cite{dupuy2015}; (127) \cite{greco2019}; (128) \cite{liu2011}; (129) \cite{gagne2015}; (130) \cite{liu2016}; (131) \cite{best2017}; (132) \cite{dupuy2012}; (133) \cite{bouy2004}; (134) \cite{radigan2013}; (135) \cite{faherty2016}; (136) \cite{leggett2014}; (137) \cite{bardalez2019}; (138) \cite{gaia2022}; (139) \cite{best2018}; (140) \cite{dahn2017}; (141) \cite{kirkpatrick2019}; (142) \cite{marocco2021}; (143) \cite{lodieu2019}; (144) \cite{meisner2020a}; (145) \cite{gaia2018}; (146) \cite{schilbach2009}; (147) \cite{vrba2004}; (148) \cite{tinney2003}; (149) \cite{sahlmann2016}; (150) \cite{zhang2021}; (151) \cite{smart2013}; (152) \cite{vanleeuwen2007}; (153) \cite{zapatero2014}; (154) \cite{manjavacas2013}; (155) \cite{dupuy2017}; (156) \cite{burgasser2004b}; (157) \cite{burgasser2016}}
\end{deluxetable*}

\end{document}